\pdfoutput=1
\documentclass[12pt]{article}
\usepackage{graphics, color,soul}
\usepackage{graphicx}
\usepackage{amssymb,mathtools}

\usepackage{hyperref}

\numberwithin{equation}{section}

\def\gsim{\, \rlap{$>$}{\lower 1.1ex\hbox{$\sim$}}\,}
\def\lsim{\, \rlap{$<$}{\lower 1.1ex\hbox{$\sim$}}\,}

\newcommand{\be}{\begin{equation}}
\newcommand{\ee}{\end{equation}}
\newcommand{\bea}{\begin{eqnarray}}
\newcommand{\eea}{\end{eqnarray}}
\newcommand{\vol}{{\cal V}}

\newcommand{\ba}{\begin{eqnarray}}
\newcommand{\ea}{\end{eqnarray}}

\newcommand{\beq}{\begin{equation}}
\newcommand{\eeq}{\end{equation}}

\newcommand{\phiUV}{\phi_{UV}}
\newcommand{\phiCDL}{\phi_{t}}
\newcommand{\VUV}{V_{UV}}
\newcommand{\MP}{M_{p}}

\newcommand{\bpf}{\tilde b_p^{(1)}}
\newcommand{\bpl}{\tilde b_p^{(2)}}

\begin{document}

\begin{titlepage}

\setcounter{page}{1} \baselineskip=15.5pt \thispagestyle{empty}

\vbox{\baselineskip14pt
%\hbox{hep-th/0000000}
}
{~~~~~~~~~~~~~~~~~~~~~~~~~~~~~~~~~~~~
~~~~~~~~~~~~~~~~~~~~~~~~~~~~ \footnotesize{SU/ITP-14/29, SLAC-PUB-16165, DESY-14-225}} \date{}

\bigskip\

\vspace{.5cm}
\begin{center}
{\fontsize{19}{36}\selectfont  \sc
Drifting oscillations in axion monodromy\\
\vspace{4mm}
}
\end{center}

\begin{center}
{\fontsize{13}{30}\selectfont Raphael Flauger,$^1$ Liam McAllister,$^{2}$ Eva Silverstein,$^{3,4,5}$ and
Alexander Westphal$^{6}$}
\end{center}

\begin{center}
\vskip 8pt

\textsl{$^1$ Department of Physics, Carnegie Mellon University, Pittsburgh, PA 15213, USA}

\vskip 7pt
\textsl{$^2$ Department of Physics, Cornell University, Ithaca, NY 14853, USA}

\vskip 7pt
\textsl{$^3$ Stanford Institute for Theoretical Physics, Stanford University, Stanford, CA 94305, USA}

\vskip 7pt
\textsl{$^4$ SLAC National Accelerator Laboratory, 2575 Sand Hill Rd., Menlo Park, CA 94025, USA}

\vskip 7pt
\textsl{$^5$ Kavli Institute for Particle Astrophysics and Cosmology, Stanford, CA 94305, USA}

\vskip 7pt
\textsl{$^6$ Deutsches Elektronen-Synchrotron DESY, Theory Group, D-22603 Hamburg, Germany}

\end{center}

\vspace{0.3cm}
\hrule \vspace{0.1cm}
{ \noindent \textbf{Abstract} \\[0.2cm]
We study the pattern of oscillations in the primordial power spectrum in axion monodromy inflation, accounting for drifts in the oscillation period that
can be important for comparing to cosmological data.
In these models the potential energy has a monomial form over a super-Planckian field range, with superimposed modulations whose size is model-dependent.
The amplitude and frequency of the modulations are set by the expectation values of moduli fields.  We show that during the course of inflation, the diminishing energy density can induce slow adjustments of the moduli, changing the modulations. We provide templates capturing the effects of drifting moduli, as well as drifts
arising in effective field theory models based on softly broken discrete shift
symmetries, and we estimate the precision required to detect a drifting period.  A non-drifting template suffices over a wide range of parameters, but for the highest frequencies of interest,
or for sufficiently strong drift, it is necessary to include parameters characterizing the change in frequency over the e-folds visible in the CMB.
We use these templates to perform a preliminary search for drifting oscillations in a part of the parameter space in the Planck nominal mission data.
\noindent }

\vspace{0.3cm}
 \hrule

%\vspace{0.6cm}
\end{titlepage}

\tableofcontents

\newpage

\baselineskip = 16pt

\section{Introduction: Phenomenology of Axion Monodromy}

Cosmic microwave background (CMB) and large-scale structure (LSS) observations are becoming sensitive enough to constrain broad classes of inflationary models.  Two important signatures for this purpose are primordial gravitational waves, imprinted in B-mode polarization, and the shapes
of the power spectrum and non-Gaussian correlators.
In this paper, we will be concerned with the search for oscillatory features in the scalar power spectrum.

In {\it{axion monodromy}} scenarios \cite{MonodromyI,MSW,FMPWX, ignoble}, a super-Planckian inflaton displacement results from repeated circuits of a sub-Planckian fundamental period.\footnote{This mechanism can be regarded as an ultraviolet completion of large-field chaotic inflation \cite{Andreichaotic}, with additional dynamics including some aspects of natural inflation \cite{Natural}.  As in those scenarios, there are interesting generalizations that involve multiple dynamical fields, with the possibility of distinct signatures \cite{multiax, Danjie}.} Gravitational waves are then detectably large,  while the tilt $n_s$ can take a range of values determined by the number of fields \cite{Danjie}\ and the reheating scenario.
Oscillatory contributions to the scalar power spectrum arise as a direct consequence of the underlying periodicity.  However, the oscillatory features have a model-dependent amplitude, and may be undetectably small: in particular,
the amplitude is exponentially suppressed in regimes where the oscillations are generated by instanton effects.   Nonetheless,  searching for oscillations is well-motivated, because a discovery would imply a very interesting additional structure in the primordial perturbations, and conversely a null result can constrain the parameter space in
a useful way.

A featureless, nearly scale-invariant $\Lambda$CDM power spectrum provides a very good fit to the data, with values of $\chi^2/{\rm d.o.f}$ typically just a few percent above unity.
However, given the large number of degrees of freedom in datasets such as {\it Planck}, the total excess $\chi^2$ is large enough that new physics in various forms could be hiding in the data, waiting to be discovered. Oscillatory features in the power spectrum provide one example, and there may also be more general theoretical sources of excess variance, as in the scenarios of \cite{DanGreen}.

Many searches for oscillations that are periodic in the canonical inflaton field  have been performed \cite{oscillationsdata}, with no unambiguous detection.
The goal of this paper is to point out that the symmetry structure and dynamics of axion monodromy allow for --- and often require ---  slow secular drift of the frequency of oscillations.   One key source of this drift is backreaction of the inflationary sector on other degrees of freedom, such as  moduli scalars in string compactifications.  Drifting oscillations can also be motivated from low energy effective field theory considerations \cite{EFToscillations}, assuming a weakly broken discrete shift symmetry.
We will analyze the conditions under
which the drift becomes large enough so that  limits set with a constant-frequency oscillatory template are inaccurate, finding that for sufficiently high frequencies or for sufficiently strong evolution of the period,  a drifting-frequency template is required.

The mechanism underlying monodromy, involving a weakly broken shift symmetry, is quite robust, and arises naturally in string theory from the couplings of gauge potential fields to branes and to fluxes.  This structure extends via string-theoretic duality relations to certain other sectors of fields as well.   However, as with all inflationary mechanisms in string theory, a completely systematic exploration of axion monodromy remains a distant goal.
A more practical approach at present --- which we will adopt in this paper --- is to determine the  observational  limits on phenomenological models whose parameter ranges are informed by  microphysical constraints that arise in string theory.\footnote{Structural constraints analogous to those used in this work can be found in \cite{SC}.}

The potential in single-field axion monodromy inflation takes the general form
\be\label{form}
V(\phi)=V_0+\mu^{4-p}\phi^p + \Lambda(\phi)^4 \cos \Bigl(\gamma_0+\phi/f(\phi) \Bigr)\,,
\ee
where $V_0$, $\mu$, $\gamma_0$, and $p$ are constants, and we have focused on the large-field regime of the potential.
The variable coupling $f$, which we will refer to as the axion decay function, reduces to the usual axion decay constant $f$ when $f$ is constant.
Examples in the literature include cases with $p=3,2,4/3,1,2/3$.

In string theory realizations of axion monodromy,  $\Lambda$ and $f$ are functions of the moduli fields, which can shift over time as the inflationary energy decreases.
As we will explain, this effect is analogous to the {\it{flattening}} of the potential that arises as a result of shifts of moduli vevs during inflation \cite{flattening,powers}.   Indeed,  drift and flattening are two natural consequences of the adiabatic evolution of massive moduli.
However, it is interesting to note that flattening, which  determines the power $p$ and the resulting slow-roll parameters, is generally independent of the drift in $f$: different degrees of freedom may adjust in the two cases, and  moreover the final value of $p$ depends on the fiducial value $p_0$, which is a separate parameter.  So the drift in frequency is not determined purely by an expansion in the slow roll parameters of the non-oscillatory potential,  even though the latter does make a model-independent contribution to the drift that is calculable within effective field theory.

In this work we will present templates that parameterize key  features of ultraviolet-complete examples of axion monodromy inflation, without being so specific as to be tied to a  particular realization.  We will provide one specialized template based on the pattern of frequency drift deduced from simplified models of axion monodromy, along with a more general Taylor series expansion  that captures more general models, including \cite{MSW,FMPWX}.
We anticipate that these templates will have significant overlap with the predictions of a broad range of models.  This is similar in spirit to the use of simplified models of physics beyond the standard model in the analysis of LHC data.

We will examine oscillations with a drifting frequency in a number of toy models in order to illustrate the  spectrum of possibilities.
We will scan over a range of
the parameters entering the templates,
including the parameter values exhibited by the models.  For models developed thus far in the literature, at most frequencies the drift is weak enough that a signal would be caught by
searching for a
non-drifting template.   However,
we do not find firm predictions of string theory for the parameter ranges, because the space of possible models is just beginning to be explored (see \cite{powers,recentmonodromy} for recent developments.)  We therefore propose significantly larger ranges for the drift parameters, in addition to scans over a wide range of frequencies, including all values for which the low energy effective field theory is under control and radiatively stable.  In some of this range, we find that the drift cannot be fully captured by a Taylor expansion of the axion period.
Ultimately, the  analysis  presented here does not lead to highly informative priors on the parameters involved in drifting oscillations.  Rather, it highlights the need to incorporate such parameters in searches for oscillatory features in part of the parameter space.  Since the number of additional parameters will not be large, searches for oscillations in cosmological data can still yield useful constraints.

The organization of this paper is as follows.   In \S\ref{generaltheory} we  discuss the causes of frequency drift and the estimate the importance of this drift in analyses of cosmological data.
In \S\ref{prototypes} we present a number of examples of axion monodromy inflation scenarios, determining the drift in frequency and amplitude, and motivating classes of templates and parameter ranges.  We then pause to discuss  the amplitude of oscillations, in \S\ref{sec:creationmyth}, and the symmetry structure of axion monodromy, in \S\ref{symmetry}.
In \S\ref{analysis} we  present the results of an initial search for drifting-frequency oscillations in the {\it{Planck}} nominal mission data.
Our conclusions appear in \S\ref{theconclusion}.

\section{Drifting Period: General Theory} \label{generaltheory}

In the single-field version of axion monodromy inflation --- or any similar mechanism exhibiting a softly broken discrete shift symmetry ---  one finds a potential for the canonically normalized inflation field $\phi$ of the form
\be\label{genpot}
V(\phi)=V_{0}(\phi) + \Lambda(\phi)^4 \cos[a(\phi)]\,.
\ee
Here $a(\phi)$ is the underlying periodic axion variable, which in general is a nonlinear function of the canonical inflaton $\phi$.
Nontrivial dependence $a(\phi)$ can have multiple underlying causes, including backreaction of the inflationary energy on compactification moduli, as well as loop effects derived from the weak explicit breaking of the discrete shift symmetry in $V_0(\phi)$.  In (\ref{genpot}) we have also allowed the   amplitude $\Lambda^4$ to depend on $\phi$, but  because amplitude drift is generally less important than frequency drift  in the search for oscillatory features,  our analysis will primarily focus on the drifting frequency encoded in $a(\phi)$.

\subsection{Drift from backreaction on string moduli}

In string-theoretic models of axion monodromy,  the leading contribution to the drift in period comes from the coupling of the axion to additional scalar fields $\sigma_I$ known as {\it{moduli}}.  The moduli fields include the size and shape of the extra dimensions of string theory, as well as the coupling $g_s$.  Axion-moduli couplings  can appear in both the kinetic and potential terms, and  their effect on the leading potential term $V_0(\phi)$ has been studied in \cite{flattening,powers}.  These effects can become very complicated in general, but much has been learned from the basic structures involved in existing moduli stabilization proposals.  (In fact, allowing the inflationary potential to participate in moduli stabilization can simplify the latter \cite{powers,SC}.)

The nontrivial dependence $a(\phi)$ in (\ref{genpot}) arises from the moduli-dependence of the axion decay function $f$ ($\ne {\rm{constant}}$) in the axion kinetic term.  Because the moduli $\sigma_I$ in general adjust during inflation as a result of their couplings to the axion potential terms, we  can write
\be\label{genkin}
\int d^4 x \sqrt{-g} f[\sigma_I(a)] \dot a^2 = \int d^4 x \sqrt{-g} \dot\phi^2\,,
\ee
with $\phi$ the canonical field, given by the solution to $\frac{d\phi}{da} = f(a)$.
The theory contains sectors that are periodic in the axion $a(\phi)$, leading to the oscillatory term in (\ref{genpot}).

The amplitude of these oscillations is highly model-dependent, and also in general depends on the moduli fields and hence on $\phi$.  Since there is a regime of parameters
in which the leading oscillations
arise from instanton effects, these can easily be exponentially suppressed as a function of the natural couplings of the theory.  However, in high-scale inflation, there are limits on how weak these couplings can be, leaving room for a detectable signal, as we explore in more detail in \S\ref{sec:creationmyth}.

\subsection{Required precision}  \label{requiredprecision}

The comparison between models with oscillatory power spectra and data is commonly based on a search for a given set of templates and constraints on their parameters~\cite{oscillationsdata}. To assess the importance of the drifting period for comparison to cosmological data we must thus understand whether a model with drifting period would have led to a detection in these searches or might have been missed. For simplicity we consider an ideal power spectrum measurement for a Gaussian random variable such as the scalar perturbation $\zeta$ or the temperature anisotropies $a_{\ell m}$. The likelihood for a theoretical power spectrum $P_k$ given a measurement $\hat{P}_k$ is
\begin{equation}
\mathcal{L}(P|\hat{P})\propto \exp\left[-\frac12\sum\limits_k N_k\left(\frac{\hat{P}_k}{P_k}+\ln P_k-\frac{N_k-2}{N_k}\ln \hat{P}_k\right)\right]\,,
\end{equation}
where $N_k$ is the number of modes contributing to $P_k$. The theoretical spectra are typically taken to be of the form $P_k=P^0_k+A\delta P_k$, where $P^0_k$ is a smooth spectrum, and $\delta P_k$ is an oscillatory template that will depend on a number of parameters such as the frequency, the phase of the oscillations, etc.  We have explicitly introduced one of them, the amplitude $A$. Oscillatory contributions to the angular power spectrum in the absence of a drift are typically constrained at the few percent level, so we consider $A\ll1$ for our present discussion. In this limit, the amplitude that maximizes the likelihood becomes
\begin{equation}
\hat{A}=\frac{{\textstyle \sum_{k}} w_k(\hat{P}_k-P^0_k)\delta P_k}{{\textstyle \sum_{k'}} w_{k'}\delta P_{k'}\delta P_{k'}}\,,
\end{equation}
where $w_k=N_k/2(P^0_k)^2$, so that the different contributions are inverse-variance weighted, and the variance for small $A$ is
\begin{equation}\label{eq:var}
\Delta A^2=\frac{1}{{\textstyle \sum_{k'}}w_{k'}\delta P_{k'}\delta P_{k'}}\,.
\end{equation}
Let us now assume that the true power spectrum contains oscillations, but of a shape different from the template $\delta P_k$, so that
\begin{equation}
\langle\hat{P}_k\rangle=P^0_k+\delta P^{\rm true}_k\,.
\end{equation}
For this power spectrum, the expected amplitude is
\begin{equation}\label{eq:Ab}
\langle \hat{A}\rangle=\frac{{\textstyle \sum_k } w_k \delta P^{\rm true}_k\delta P_k}{ {\textstyle \sum_{k'} } w_{k'}\delta P_{k'}\delta P_{k'}}\,,
\end{equation}
Using equations~\eqref{eq:var} and~\eqref{eq:Ab}, and introducing the notation $\delta P\cdot\delta P'=\sum\limits_k w_k\delta P_{k}\delta P'_{k}$, we can write the signal-to-noise ratio for our template as
\begin{equation}
\frac{S}{N}=\frac{\delta P\cdot\delta P^{\rm true}}{\sqrt{\delta P\cdot\delta P}}=\frac{\delta P\cdot\delta P^{\rm true}}{\sqrt{\delta P\cdot\delta P}\sqrt{\delta P^{\rm true}\cdot\delta P^{\rm true}}} \sqrt{\delta P^{\rm true}\cdot\delta P^{\rm true}}\,.
\end{equation}
Thus, the signal-to-noise ratio we expect for a given template is the product of the signal-to-noise ratio for the correct template times the overlap, or cosine, between the two templates
\begin{equation}
\left(\frac{S}{N}\right)_{\rm template}=\cos(\delta P,\delta P^{\rm true})\times\left(\frac{S}{N}\right)_{\rm true}\quad \text{with}\quad \cos(\delta P_1,\delta P_2)=\frac{\delta P_1\cdot\delta P_2}{\sqrt{\delta P_1\cdot\delta P_1}\sqrt{\delta P_2\cdot\delta P_2}}\,.
\end{equation}
 In the approximation in~\cite{FMPWX}, the power spectrum for the potential~\eqref{genpot} is
\begin{equation}\label{eq:powspec}
P(k)=P(k_\star)\left(\frac{k}{k_\star}\right)^{n_s-1}\left(1+\delta n_s(\phi)\cos[a(\phi_k)]\right)\,,
\end{equation}
with $\phi_k$ the value of the scalar field at the time the mode with comoving momentum $k$ exits the horizon. For a CMB power spectrum measurement the number of modes contributing to a given multipole is $2\ell+1$. To define an inner product on primordial power spectra that approximates the inner product on angular power spectra, we choose $N_k=k$. For a power spectrum of the form~\eqref{eq:powspec}, and ignoring the slow variation of the amplitude, the inner product is
\begin{equation}
\delta P_i\cdot \delta P_j\propto\int\limits_{k_\text{min}}^{k_\text{max}} k\, dk \cos[a_i(\phi_k)] \cos[a_j(\phi_k)]\,,
\end{equation}
To estimate the level of precision we should require of our templates, let us write\footnote{This definition of $f(\phi)$ will be convenient, even though it differs from the definition obtained from the kinetic term by solving $f da = d\phi$.} $a(\phi)\sim \phi/f(\phi)$ and perform a Taylor expansion of $f$ about some point $\phi_0$ in field space.
\be\label{Texp}
\cos[a(\phi)]=\cos\Biggl[\frac{\phi}{f_0} \times \Biggl(1+\frac{\phi_{0}}{f_0}\frac{df}{d\phi}\Bigr|_{\phi_{0}}\left(\frac{\phi-\phi_{0}}{\phi_{0}}\right)+\frac{1}{2}\frac{\phi_{0}^2}{f_0}\frac{d^2f}{ d\phi^2}\Bigr|_{\phi_{0}}\left(\frac{\phi-\phi_{0}}{\phi_{0}}\right)^2+\dots\Biggr)^{-1}\Biggr]\,.
\ee
In axion monodromy scenarios in which $V_0(\phi)$  is approximately monomial, $V_0(\phi) \approx \mu^{4-p}\phi^p$, we find that $\phi_{0}$ is of order $10 M_p$ when the modes observed in the CMB exit the horizon, and the change in the field is $\Delta\phi_{\rm CMB}\sim M_p$ during this time.\footnote{Although we will focus on the single-field version of axion monodromy for simplicity in the present work, the multifield version of the mechanism \cite{multiax,Danjie} involves smaller individual field ranges.}
The expansion (\ref{Texp}) is therefore organized as a power series in $|\phi-\phi_{0}|/\phi_{0} \lesssim 1/10$
with coefficients $a_n= \frac{1}{n!}\phi_{0}^n f_0^{-1}\frac{d^n f}{d\phi^n}|_{\phi_{0}}$.  We will find below that based on simple estimates of the effect of backreaction on moduli, the coefficients $a_n$ can easily be of order  unity.  For example,
we will see that there are models in which
\be\label{powerfcase}
f(\phi)=f_0\left(\frac{\phi}{\phi_0}\right)^{-p_f}\,,
\ee
with a power $p_f$ of order unity.

Under the assumptions made in~\cite{FMPWX}, the oscillatory contribution to the power spectrum is obtained from equation~\eqref{Texp} by replacing $\phi$ by $\phi_k\approx\sqrt{2p(N_\star-\ln k/k_\star)}$. We will discuss in \S\ref{sec:srcorr} to what extent these approximations are valid, but for now we use them to illustrate the comparison between drifting and non-drifting models using the inner product.
One should keep in mind that searches
for oscillatory features in the data vary the phase and frequency of the oscillations. To assess whether a signal with a drifting period might have been missed in previous searches, we should thus maximize the overlap between the templates with fixed and drifting frequency by allowing the parameters of the fixed-frequency template to vary. Setting $k_\text{min}=10^{-4} {\rm Mpc}^{-1}$, $k_\text{max}=10^{-1} {\rm Mpc}^{-1}$, and $k_\star=0.05 {\rm Mpc}^{-1}$, we find that for a frequency close to the WMAP9 best-fit value $f/M_p\sim 4\times 10^{-4}$, overlaps as large as $80\%$ can be achieved between the template for constant axion decay function and the template with the coefficient of the first correction
being
of order unity. However, the overlap drops rapidly as the frequency of the oscillations increases. For $f/M_p\sim 2\times 10^{-4}$ we find overlaps of around $50\%$, and for the lower end of the range of axion decay constants studied here, $f/M_p\sim 10^{-4}$, the overlap is further reduced to only a few percent. We conclude that at least two nontrivial terms in the Taylor expansion (\ref{Texp}) of the axion decay function  are necessary to achieve an order unity overlap for frequencies $f/M_p\lesssim 2\times 10^{-4}$. Consequently, previous searches do not cover the entire space of models with drifting frequencies.

We should note that there is not always a useful description in terms of a Taylor expansion. For example,  this expansion breaks down for large $p_f$.  This is another regime in which one cannot rely on a non-drifting template.
With such a strong drift in $f$, one would need to estimate radiative  corrections to the slow roll parameters,
which we estimate in \S\ref{sec:radcorr}.  For a small enough amplitude of modulations, such corrections can be neglected as compared to $V_0(\phi)$, but at some point the oscillations would then become too small to be detectable.

\subsubsection{Slow roll corrections to the oscillatory power spectrum}\label{sec:srcorr}

The derivation in \cite{FMPWX} used above is only accurate to leading order in
the
slow roll parameters.  Given the monomial large-field form  of the leading potential $V_0(\phi)\sim \mu^{4-p}\phi^p$, the slow roll parameters $\eta$ and $\epsilon$ are of order $(M_p/\phi)^2\sim 10^{-2}$.  Let us estimate their importance in the power spectrum by considering corrections to the argument $a(\phi)$ of the cosine of the form
\be\label{SRexp}
\cos\left[\frac{\phi_k}{f}\left(1+\sum c_n\left(\frac{M_p}{\phi_k}\right)^{2n}\right)\right]\,.
\ee
Evaluating the overlaps between templates with coefficients $c_n$ of order unity suggests that both the $n=1$ and $n=2$ terms in this expansion are important.
This should not come as a surprise. In order to obtain a good approximation to the oscillatory features in the power spectrum,  one must keep correction terms in  $\Delta a\sim \frac{\phi_k}{f}\sum c_n\left(\frac{M_p}{\phi_k}\right)^{2n}$
that can be of order $2\pi$ or bigger.

Extending the analytical derivation by two orders in the slow-roll expansion is beyond the scope of this paper. For our numerical analyses, we will therefore check that a given template agrees with a numerical calculation.

\subsection{Radiative corrections and strong drift}\label{sec:radcorr}

To estimate whether the low-energy effective theory is radiatively stable even when the Taylor expansion breaks down, let us consider the case (\ref{powerfcase}) with $|p_f| \sim {\cal{O}}(10)$.
We change variables back to the axion $a=a_0(\phi/\phi_0)^{1+p_f}$, with $d\phi = \hat f(a) da \Rightarrow \hat f(a_0) \equiv \hat f_0 = \phi_0/a_0(1+p_f)$.
Starting from a typical axion monodromy potential $V_0(\phi)\approx \mu^{4-p}\phi^p$, we can write
\bea\label{Vforms}
{\cal L} &=& \frac{1}{2}(\partial\phi)^2- \mu^{4-p} \phi^p -\Lambda_0^4\cos\left[\gamma_0 +\frac{\phi_0}{f_0} \left(\frac{\phi}{\phi_0}\right)^{p_f+1}\right] \\  \label{aform}
&=& \frac{1}{2}\left(\frac{a}{a_0}\right)^{-2 p_f/(1+p_f)} (\hat f(a_0)\partial a)^2 - \hat\mu^4 a^{p/(p_f+1)}-\Lambda_0^4\cos[a]\,,
\eea
where for simplicity we here neglect any drift in the amplitude $\Lambda^4$.

Each form of the action makes manifest a different limiting symmetry.  As $\mu\to 0$ and $\Lambda_0^4\to 0$,
the first form (\ref{Vforms}) exhibits a continuous shift symmetry in $\phi$.  The second form (\ref{aform}) makes clear that the oscillatory term by itself respects the discrete shift symmetry $a\to a+2\pi$.  These symmetries constrain the form of the corrections.

One can estimate corrections using the first form of the action, starting from the one-loop Coleman-Weinberg effective potential.  Taking into account the leading contributions from this, the essential effect of $p_f$ is to generalize the expansion in $\Lambda_0/(4\pi f)$ to an expansion in $(p_f+1)\Lambda_{0}/(4\pi f)$.
We will now explore this using the second form of the action (\ref{aform}).
 
Our basic question is whether the strong drift parameterized by $p_f\sim 10$ generates corrections
that ruin slow roll inflation.  To assess this, we can focus on the interactions coming from the kinetic plus oscillatory terms, which we will now delineate.

In the expansion of the kinetic term, the Lagrangian (\ref{aform}) exhibits a sequence of higher dimension operators, with coefficients that depend on our parameter of interest, $p_f$.  Expanding this (defining $a=a_0+\delta a$), we find
\be\label{aexpansion}
\frac{1}{2}(\partial[\hat f(a_0)\delta a])^2\left(1-\hat f(a_0)\delta a\left\{\frac{1}{\hat f(a_0) a_0}\left(\frac{2p_f}{p_f+1}\right) \right\}+\dots\right)\,.
\ee
The dimension-one fluctuating field in this Lagrangian is $Y\equiv \hat f(a_0)\delta a$.
The higher-dimension operators arising from this expansion of the kinetic term are suppressed by
the mass scale
\begin{equation}
M \equiv |\hat f(a_0)a_0|=\phi_0/|1+p_f| \sim \frac{10}{|1+p_f|} M_p\,.
\end{equation}
If for example $p_f\sim 10$, then this  suppression scale is still very high, $M \sim M_p$.
In particular, it is much higher than the strong coupling scale $\sim 4\pi \hat f_0$ of the low energy theory.
We will assume that the low energy theory is effectively cut off at a scale $\Lambda_{UV}\ll \hat f$,
and estimate the radiative stability of slow roll inflation against the loop corrections generated by the interactions in (\ref{Vforms}), with the loop momenta cut off at
$\Lambda_{UV}$.  It would be interesting to analyze this in the context of field theoretic (as well as string-theoretic)
ultraviolet completions that describe the degrees of freedom at and above the scale $4\pi\hat f$.

This latter scale is evident if we expand the oscillatory term:
\be\label{cosexp}
\Lambda_0^4 \cos(a) =\Lambda_0^4\cos\left(a_0+\frac{Y}{\hat f_0}\right)= \Lambda_0^4\sum
c_j\left(\frac{Y}{\hat f_0}\right)^{j}\,.
\ee
However, note that the background solution for $Y$ ranges over a distance in field space much greater than the underlying period $\hat f_0$, so one cannot use the last form expanded about $Y=0$ for the whole process.  In the original cosine form, it is clear that the magnitude of the interaction term is bounded by $\Lambda_0^4$.
(Consider the one-loop Coleman-Weinberg potential ${\rm Tr}[\log (\partial^2+V''(\varphi_0))]$ in scalar field theory:  for our sinusoidal potential $V$, one has
$V''= -(\Lambda_0^4/\hat f^2)\cos(a_0+Y/\hat f_0)$, which gives smaller corrections than would be estimated
from the individual terms in the series (\ref{cosexp}).)

Let us first estimate an upper bound on the corrections to the potential term $V(a)$ and their effect on the slow roll parameters, assuming for this discussion that the kinetic term is not corrected in an important way.  As we just noted, by symmetry the only potentially relevant corrections must use a combination of the interaction vertices from the kinetic term and the oscillatory term.
Some corrections contribute to the oscillatory term in the potential, proportional to $\cos(a)$ and $\sin (a)$ (with mildly drifting amplitude); we find these are controllably small.   Others are proportional to even powers of $\cos( a)$ and $\sin( a)$, and introduce corrections to the non-oscillatory potential $V_0(\phi)$.  The corresponding leading contribution to the slow roll parameter
$\epsilon_V=\frac{M_p^2}{2}(V'/V)^2$ arises from
\be\label{DeltaVprime}
\Delta V' = \frac{dV}{d\phi}=\frac{\hat f_0}{\hat f}\frac{dV}{dY} \le \frac{\hat f_0}{\hat f}\frac{1}{M}\frac{\Lambda_0^8}{\hat f^4}\,.
\ee
Here in the second and third factors we introduced a vertex from the kinetic term and two from the cosine term, with the understanding that higher loop contributions will be suppressed by additional powers of $\Lambda_{UV}/\hat f$ and $\Lambda_{UV}/M$.  As discussed above, $\Lambda_{UV}$ is the scale at which the loop is effectively cut off.   We write
(\ref{DeltaVprime}) as an inequality because there may be additional cancellations
that we are not working out explicitly.
The inequality (\ref{DeltaVprime}) translates into
\be\label{Deltaepsilon}
\Delta\epsilon_V < \left( \frac{\Lambda_0^{8}}{V_0^2}\right) \left(\frac{\Lambda_0^{8}}{\hat f_{min}^8 }\right) \left( \frac{M_p}{M}\right)^2\left(\frac{\hat f_0}{\hat f_{min}}\right)^2
\ee
where we used the fact that $d\phi/da = \hat f$.  To get the most conservative bound, we take $\hat f$ at its minimum value, $\hat f_{min}$, within the range of field visible in the CMB.  The factor
$(M_p/M)^2$ is of order 1.  The last factor is $\lesssim 1$, as the series of higher order terms should resum into a cosine-type dependence in each diagram.  Using $\hat f\propto (\phi/\phi_0)^{-p_f}$, we find that the
penultimate factor is $\sim (5/6)^{p_f}\lesssim 6$ (taking the range to be roughly  between 50 and 60 e-folds).  Altogether this does not put a stringent bound on $\Lambda_0^4/V_0$.

Next, let us estimate the corrections to the kinetic term.  Consider the 1PI 1-loop diagram with a three-point vertex from the cosine term and a four-point vertex from the kinetic term.  This gives a correction to the $(Y/M)(\partial Y)^2$ term in the Lagrangian (with one $\partial Y$ leg from the cosine-term vertex and the other $Y(\partial Y)$ from the four-point vertex).  This diagram is of order $(1/M^2)(\Lambda_0^4/\hat f^3)$.  So using $M\sim M_p$, it is down from the leading term of this form $\sim (Y/M)(\partial Y)^2$ from (\ref{aexpansion}) by a factor of
$\Lambda_0^4/M_p^4(M_p/\hat f)^3$.   Now since $f/M_p$ can range down to $10^{-4}$, this correction becomes
important only if $\Lambda_0^4 \gtrsim 10^{-12}M_p^4 $, which exceeds the existing observational upper bound (\ref{experimentalbound}).
Again, higher loops will have extra suppression by powers of $\Lambda_{UV}/\hat f$ and $\Lambda_{UV}/M$.  Diagrams with more external legs from the vertices obtained from the expansion of the cosine term (\ref{cosexp}) will only contribute terms $\sim Y^n(\partial Y)^2$ over a small range $\ll \hat f$ in $Y$.

As a result, there appears to be a regime of parameters in which the Taylor series template fails, but the shift-symmetry breaking from the drift is not so large as to produce substantial corrections to the underlying inflationary slow roll mechanism.   This regime is the most challenging one for the problem of searching for oscillations.  It requires knowing the functional form of the drift in period, without the benefit of a convergent Taylor expansion.   The example we just gave illustrates this possibility.

\section{Drifting Period: Prototypes}  \label{prototypes}

In this section we explore the form of the function $a(\phi)$ in  a few classes of  prototypical examples of moduli stabilization and axion monodromy inflation in string theory.
In \S\ref{sec:powerlaw} we examine `power-law' models in which classical sources of stress-energy lead to polynomial dependence on the moduli fields.
Then, in \S\ref{sec:nonperturbative} we turn to scenarios in which nonperturbative quantum effects play a central role in stabilization.

Let us start by briefly reviewing the general structure of axions and their decay functions $f$.  Axions $a_I$ arise from  higher dimensional analogues of electromagnetic potential fields $A_{(q)}$,
\be\label{aA}
A_{(q)}=\sum a_I\,\omega_{(q)}^I\,,
\ee
where the $\omega_{(q)}^I$ are a basis for the cohomology $H^q(X)$  of the compactification manifold $X$.
There are various duals or analogues of axions under electromagnetic and target space string dualities, in some of which the axion becomes a brane collective coordinate or a geometric modulus.  In fact, the possibility of monodromy inflation was discovered in the context of such examples \cite{MonodromyI} (cf.~\cite{twistedtori}),  and interesting mechanisms of that kind continue to appear \cite{unwinding}.  These duals have similar drifts in their oscillatory periods, so for simplicity we will focus on axions that descend directly from higher dimensional potential fields.

The kinetic term $\int |dA|^2$ for $A$ descends to kinetic terms for $a$, which generally depend on the moduli fields as indicated in (\ref{genkin}).   The potential descends from gauge-invariant terms in the $D$-dimensional effective action of the schematic form
\be\label{Stuckelberg}
\int d^D x\sqrt{-G}\sum_q|\tilde F_q|^2 =\int d^D x\sqrt{-G}\sum_q|F_q-C_{q-3}\wedge H_3 +\dots + F_{\tilde q} B\wedge \dots\wedge B|^2\,,
\ee
in terms of potential fields $A=\{C_n, B_2\}$ and their field strengths ${dA}=\{F_{n+1}, H_3\}$.  The combinations $\tilde F_q$ are known as generalized fluxes.
The couplings in (\ref{Stuckelberg}) generalize the Stueckelberg term $\int |\partial\theta -A|^2$ in spontaneously broken electromagnetism.
In the presence of generic background fluxes, these terms contain direct dependence on $A$, which descends to a monodromy-unwound potential for $a_I$.  Just as in the kinetic terms, these potential terms generally depend on the moduli $\sigma_I$.
Correspondingly, the  potential energy $V(a)$ of the axion fields $a_I$ must be incorporated in a consistent way in the metastabilization of the moduli: either $V(a)$ must be subdominant to the remainder of the moduli potential $V_{mod}$, or else  the forces encoded in $V(a)$ must be balanced against additional contributions in $V_{mod}$.
The latter case --- which may be more generic because it requires less of a hierarchy of scales in the problem --- leads to  significant drift in the oscillation period $f(\phi)$, as well as flattening of the potential \cite{flattening,powers}.   In this case the role of the axion potential energy in the overall scalar potential is analogous to the contribution of an ordinary flux.
These features have been reviewed recently in \cite{powers,LesHouches}.

In some circumstances, such as in the early examples \cite{MSW, FMPWX}, the most direct description of the monodromy is that it is induced by a brane's DBI action.
One can use the gravity solution for the brane, as in AdS/CFT, to relate the brane and flux pictures \cite{flattening}.
We will consider both cases in our explorations below.

\subsection{Power law stabilization}  \label{sec:powerlaw}

We begin by discussing scenarios in which moduli such as the volume ${\cal V}$,  the string coupling $g_s$, and  the axion decay function $f$ are stabilized by power-law effects.
By this we mean that the forces fixing these fields are powers of ${\cal V}$, $g_s$,  and $f$,  such as those arising from classical stress-energy sources (including curvature) in string compactifications.\footnote{See \cite{LesHouches}\ and references therein for pedagogical introductions to this approach to moduli stabilization.}
This type of model was studied recently in \cite{powers}, and here we extract some lessons  of that work for the  question of drifting oscillations.   In other scenarios, such as \cite{KKLT} and \cite{LARGE},  moduli stabilization results from a competition between perturbative and nonperturbative effects.  We explore that case in \S\ref{sec:nonperturbative}\ below.

One general feature of string compactifications is that the four-dimensional effective potential  vanishes in  the limit of weak coupling  and in the limit of infinite volume of the extra dimensions.  In these directions in field space, metastabilizing moduli requires (at least) three terms,  with a structure like
\be\label{threeterm}
V= a_V x^{\gamma_a}-b_V x^{\gamma_b}+ c_V x^{\gamma_c}\,,
\ee
with $x\sim x_0 e^{c_x\sigma_x/M_p}$,  where $\sigma_x$ is the canonically normalized field and $c_x$ a constant.
Metastability requires  a sufficiently strong intermediate negative term; for example for the simple case $\gamma_a=2,\gamma_b=3, \gamma_c=4$ we require
\be\label{abc}
1< \frac{4ac}{b^2}<\frac{9}{8}
\ee
for a metastable minimum.

In other directions, such as those describing flux-stabilization of the relative sizes of dual cycles in the manifold, one finds a simpler two-term structure of the schematic form
\be\label{twoterm}
V=A_V y^{\delta_A}+\frac{B_V}{y^{\delta_B}}\,,
\ee
with again $y\sim y_0e^{c_y\sigma_y/M_p}$.  The prototype for this is a manifold with a product structure $M_1\times M_2$ with flux threading the two orthogonal directions $M_1$ and $M_2$:  in each direction, the flux prevents collapse of the cycle, leading to a potential of the form (\ref{twoterm}) for the ratio of the cycle sizes.

In general, the coefficients $a_V, \dots, B_V$ can depend directly on axions, depending on the details of the example.  In particular, we can see immediately from the terms (\ref{Stuckelberg}) that the role of fluxes is more generally played by combinations of fluxes and axions.  Since at large field values (or with many axion fields \cite{multiax}), the axion potential in itself satisfies the slow roll conditions, it can be consistent for the axion-dominated terms to participate in moduli stabilization \cite{SC,powers}.

In three-term stabilization, however, there is a significant limitation on the contributions of axions to the scalar potential.  For a single axion  that evolves over a range of order $10 M_p$ --- implying many  circuits of its underlying period --- the condition (\ref{abc}) can easily fail by virtue of too large a change in $4ac/b^2$ over this large field range.\footnote{If many axions each move over a very small range, this problem can be ameliorated.}
In two-term stabilization,  in contrast, it is straightforward for axions to participate much like fluxes.

\subsubsection{Motivating examples}  \label{motivatingexample}

Let us next describe an illustrative pair of examples drawn from \cite{powers}, before attempting to draw some more general lessons that will motivate  the oscillatory templates and parameter ranges presented in this work.  The drift in oscillations arises in these scenarios after stabilizing one combination of the string coupling $g_s$ and internal volume ${\cal V}$ with a three-term structure, and another combination with a leading contribution from the axion $b$ driving inflation.
Let us work from equation (4.8) of \cite{powers}, giving the effective potential after three-term stabilization of a combination $x=g_s/{\cal V}^{2/3}$ (with $\vol$ the volume in string units)
\be\label{UnewIIi}
{\cal U}|_{x=x_{min}} \sim M_p^4\left\{ C_h^2 n_3^2 \frac{1}{\vol^{2/3}}+C_h^4 (q_3^2+q_1^2b^2) \vol^{2/3} + C_h^4 (q_5^2 + 2 q_5 q_1 b^2 + q_1^2 b^4)\right\}\,,
\ee
where $C_h$ is a constant depending on  the parameters of the underlying model, $n_3$  is a quantum number corresponding to $H_3$ flux, and the $q_i$  similarly correspond to Ramond-Ramond fluxes, as described in \cite{powers}.  This  potential is
valid as long as $2 q_5 q_1 b^2 + q_1^2 b^4\le {\cal O}(q_5^2)$.

The first two terms here stabilize the volume $\vol$.  In the case where $q_1^2 b^2 \gg q_3^2$, we obtain
\be\label{bvol}
b\propto 1/\vol^{2/3}
\ee
from this.  The relation (\ref{bvol}) has two effects:  first, it introduces a linear term in $b$ in the potential,  after substituting (\ref{bvol}) in (\ref{UnewIIi}).
Second,  it changes the relation between $b$ and the canonically normalized inflaton field $\phi_b$ because of the $\vol$ dependence in the $b$ kinetic term:
\be\label{kincan}
\frac{\dot b^2}{\vol^{2/3}}\propto \dot b^2 b \Rightarrow \phi_b\propto b^{3/2}\,.
\ee
This implies a drift in $f$ given by
\be\label{fdriftexamples}
f\propto \frac{1}{\vol^{1/3}}\propto b^{1/2}\propto \phi^{1/3}\,.
\ee
Therefore the oscillatory term in the potential takes the form
\be\label{oscpowerI}
\Lambda(\phi)^4\cos\left[ \gamma_0+ \frac{\phi_0}{f_0}\left(\frac{\phi}{\phi_0}\right)^{1+p_f}\right], ~~~~ p_f=-1/3\,.
\ee
where $\phi_0$ is the field value at a pivot scale, e.g.~at the start of the observable stage of inflation.
In the framework \cite{powers} in which (\ref{UnewIIi}) applies, the leading term $V_0(\phi)$ is of the form
\be\label{psexamples}
V_0(\phi)=\mu^{4-p}\phi^p, ~~~~ p=4/3\quad {\rm{or}}\quad p=2/3\,,
\ee
with $p=4/3$ arising when the term
$q_1q_5 b^2$ dominates in the potential, and $p=2/3$ arising when the term $q_1^2 b^2\vol^{2/3}$ dominates in the potential.

The amplitude of oscillations $\Lambda(\phi)^4$ is a very model-dependent quantity, one that as emphasized above can easily be undetectably small.  In the examples defined by (\ref{oscpowerI}), (\ref{psexamples}), we can estimate their size as given by worldsheet instanton effects \cite{worldsheetinstantons}\ derived from processes  in which the string wraps a two-cycle of the internal manifold.   Contributions to the potential from worldsheet instantons are exponentially suppressed in $\vol^{1/3}$, suggesting an amplitude
\be\label{Lambdaexamples}
\Lambda(\phi)^4 = \Lambda_0^4 e^{-\vol_0^{1/3}\left(\frac{\phi}{\phi_0}\right)^{p_\Lambda}}, ~~~~ p_\Lambda=-1/3\,.
\ee
where $\vol_0$ is the volume in string units  evaluated at the pivot scale.
The  amplitude (\ref{Lambdaexamples}) is exponentially suppressed in an internal two-cycle volume, something sufficiently model-dependent that it prevents us from obtaining a model-independent signature from the oscillations in axion monodromy.   But for high-scale inflation, the size of the extra dimensions is naturally at the inverse GUT scale or below, and the amplitude depends on details such as flux quantum numbers and topology.  There is a range of amplitudes of interest, including both a regime where the effect is large enough to be detectable as well as a regime where it is too suppressed to be seen.  In section \S\ref{sec:creationmyth}\ below, we will discuss the amplitude in slightly more detail, and also explore a scenario for initial conditions in which the amplitude is bounded from below.

\subsubsection{More general powers in single-scale models}

The examples just discussed motivate a structure for the inflaton potential of the form
\be\label{powersform}
V\approx \mu^{4-p} \phi^p +\Lambda_0^4\, e^{- C_0 \left(\frac{\phi}{\phi_0}\right)^{p_\Lambda}}\cos\left[\gamma_0 +\frac{\phi_0}{f_0} \left(\frac{\phi}{\phi_0}\right)^{p_f+1}\right]\,,
\ee
with powers $p, p_\Lambda, p_f$  that may take more general values than those found in the above scenario.
To search for drifting oscillations in cosmological data, we also require an understanding of the range over which these parameters can vary.
The drift in frequency arising from $p_f$ is more important to keep track of than the amplitude (related to $p_\Lambda, C_0$ here) or the underlying slow-roll potential (related to $p$), and we will on a first pass set $C_0=0$ and $p_\Lambda=0$.  The allowed range of axion decay constants $f_0$ will be discussed in the next subsection.
To get an
idea of the potential range of the parameter $p_f$, let us focus on  the example of an axion $b$ descending from the Neveu-Schwarz two-form potential $B$  that couples to the fundamental string.  In a simple model with one length scale $L$ (in units of the string tension), the kinetic term $\int |dB|^2$ descends to a kinetic term of the form (\ref{genkin}) with
\be\label{fB}
\frac{f}{M_p}\sim \frac{1}{L^2}\,,
\ee
as is reviewed for example in \cite{LesHouches}.

As in the above examples, the axion potential energy may contribute to stabilization of $L$.
In a  power-law mechanism for moduli stabilization based on the Stueckelberg-like terms (\ref{Stuckelberg}),  and/or on forces from branes affecting the size $L$, the result is a relation of the form
\be\label{pL}
L=L_0 b^{p_{\beta}}\,.
\ee
From this we derive the relation with the canonical field $\phi$
\be\label{phib}
 f db = d\phi \Rightarrow b=(L_0^2(1-2p_{\beta})\phi)^{1/(1-2p_{\beta})}\,.
\ee
This translates into
\be\label{pttwo}
p_f = \frac{2 p_{\beta}}{1-2 p_{\beta}}\,,
\ee
and we would like to understand the range of values this parameter might take in more general models realizing axion monodromy inflation.

One basic question is whether only one sign of $p_f$ is possible.   We will show below that both signs are possible, with simple mechanisms for each case.  In what we will describe next, the two signs arise in somewhat different versions of monodromy inflation.  As discussed above, although flux-axion couplings can be
understood as the source of the monodromy effect for axions descending from potential fields in string theory, in some cases the most effective description is through a brane action, while in others the description in terms of fluxes is simplest.  In the following, we will find scenarios with $p_f>0$ from the former, and $p_f<0$ from the latter (one example being the case with $p_f=-1/3$ just discussed in \S\ref{motivatingexample}).

\noindent{\it Brane-induced monodromy}

Let us start by noting that power law stabilization with brane-induced monodromy immediately gives $p_f>0$.  The simplest way to see this is to work in a duality frame in which we have a pair of branes moving relative to each other with collective coordinate $x=\tilde L\theta$ around a circle of size $\tilde L\sqrt{\alpha'}$, with another brane stretched between them (for example, D4-branes stretching between NS5-branes).  This forms a configuration like a windup toy (see e.g.~figure 1 in \cite{MSW}).  As the toy unwinds, going to smaller values of $\phi$, the circle grows larger  because it is less constricted by the brane, and hence the period $f$ grows.\footnote{We will likewise find a growing period in the example of axion monodromy inflation induced by an NS5-brane pair in a nonperturbatively stabilized compactification: see \S\ref{KKLTexample}.}  Thus $f'<0$, corresponding to $p_f>0$  upon matching to the template (\ref{powersform}).

We can see this more specifically as follows.  The kinetic term from the NS5-brane action is proportional to $M_p^2 \tilde L \dot\theta^2$, and the potential term is of the form $\theta\tilde L C_1 + C_2/\tilde L^\gamma$, where we hypothesize a second term that stabilizes $\tilde L$ in combination with the brane system.  This gives an axion decay function $f\propto \tilde L^{1/2}$.  We have $\tilde L\propto \theta^{-1/\gamma+1}$.  Changing to canonical variables we find
\be\label{branecircle}
V_0\propto \phi^{2\gamma/(2\gamma+1)}\,, ~~~~ f\propto \phi^{-2/(2\gamma+1)}\,.
\ee
In other words, this type of toy example gives $p_f = 2/(2\gamma+1)>0$.

\noindent{\it Flux-induced monodromy}

Next, let us explore the behavior of $p_f$ in more general  models in which fluxes induce monodromy.
A reader interested only in a broad view could skip this discussion, as it mainly serves to motivate a range of values of $p_f$ in a way that generalizes the example with $p_f=-1/3$.

The case $p_{\beta}=1/2$ would correspond to $b\sim e^{\phi/M_p}$ (up to  factors of order unity), and does not give rise to an inflationary model of interest for the present discussion.
Excluding this case, there are three possible regimes:
\begin{center}
\begin{minipage}{3.5in}
\noindent{(1)} ~~ $p_{\beta}<0 ~~~  -1< p_f < 0 ~~~ {\rm kinetic ~ flattening}$\\[-.1cm]

\noindent{(2)} ~~ $p_{\beta}=0 ~~~  p_f = 0 \hskip1.15cm ~~~~ {\rm constant}~f$\\[-.1cm]

\noindent{(3)} ~~ $p_{\beta}> 0 ~~~   ~~~ \hskip 2.45cm{\rm kinetic ~ steepening}$\\[-.1cm]

\qquad(3a)  $0< p_{\beta}<1/2 ~~~ 0 <  p_f <\infty~~~ p>0 $\\[-.1cm]

\qquad (3b)  $p_{\beta}>1/2 ~~~ p_f<-1 ~~~ p<0 ~~ V\propto 1/\phi^{|p|} $\\[-.1cm]
\end{minipage}
\end{center}

\noindent Here `kinetic flattening' and `kinetic steepening' refer to the effect of the transformation (\ref{phib}) on the power $p$ in $V_0(\phi)=\mu^{4-p}\phi^p$.

The examples in the previous section had $p_f=-1/3$, falling into class (1).  More generally,  as explained in \S3.2 of \cite{flattening}, for axions on small localized cycles --- i.e.~in examples going beyond a single scale --- one also finds kinetic flattening.
Class (2) arises in the case of sufficiently rigid moduli stabilization.

Note that case (3b) implies an inverse relation between $\phi$ and $b$ (\ref{phib}).  This would lead to $p<0$, making the potential proportional to a negative power of the inflaton field.  We will call this possibility `kinetic inversion'.  This is not viable phenomenologically\footnote{Although it is not viable for early universe inflation, this case may be interesting if it can provide a novel mechanism for generating accelerated expansion in string theory. } according to current constraints on the tensor to scalar ratio $r$ and the tilt $n_s$ of the  scalar power spectrum.  This provides another illustration of the importance of taking into account the backreaction on moduli as a function of the inflaton vev.

From the relation (\ref{pL}) one might think that kinetic inversion or steepening  (i.e., $p_{\beta}>0$,  regardless of whether $p_{\beta}<1/2$ or $p_\beta>1/2$) might readily arise in some classes of perturbative stabilization, based on the idea that a larger inflationary energy creates a larger decompactification tadpole.  However, there are two problems with this argument.
First, as we saw in the example (\ref{UnewIIi}) (\ref{bvol}), after stabilizing an appropriate combination
\be\label{xgen}
x = \frac{g_s}{L^{K/2}}
\ee
via other sources in the moduli potential, it is not the case that the inflationary energy from the axion term must push toward decompactification.  In fact, in the above examples we saw that  the axion energy can push the volume $\vol\sim L^{D-4}$ toward smaller values once $x=g_s/L^4$ is stabilized in its minimum.
In addition, the relation between $b$ and $L$  is not pure monomial when $b$ is a subleading effect in the stabilization, which is an important distinction.  In particular, in three-term stabilization mechanisms (at least without large relative powers in the three terms), a single large-field axion cannot dominate, as we explained above.

The remaining possibility for phenomenologically viable inflation with kinetic steepening is case (3a).   However, in the single-scale scenarios described here,  known mechanisms for perturbative stabilization do not appear to generate case (3a).
Recall, as briefly reviewed above, that fluxes --- and their generalizations $\tilde F_q$ depending directly on axion fields (\ref{Stuckelberg}) --- can stabilize pairs of dual cycles.
However, if we were to balance Ramond-Ramond generalized flux terms in stabilizing $L$,  there is a simple restriction.  Within a generalized flux, the two-form potential field $B$ acts as a flux of rank 2 on the two-cycle $\Sigma_2$ that it threads.  To balance  this against some other generalized RR flux term, the latter has to be of lower rank on $\Sigma_2$.  Balancing the two terms then gives $b^n\sim L^{2n-\Delta}$, with $\Delta\ge 0$.  Turning this around, we can write $L\propto b^{n/(2n-\Delta)}$, implying  that $p_{\beta}$ as defined above in (\ref{pL}) satisfies $p_{\beta}\ge 1/2$.  As a result, stabilization based on balancing the energies in generalized fluxes does not populate the range $0<p_{\beta}<1/2$, but could  in principle give an example of the sort (3b), i.e.~with $p<0$, which would be excluded by constraints on the primordial perturbations.

Alternatively, let us consider the possibility that the RR generalized flux instead balances against another type of term, for example the term descending from the magnetic flux of the internal $B$ field $|dB|^2=H_3^2$,  or  the potential energy descending from the curvature of the extra dimensions.   With some combination (\ref{xgen}) stabilized in a three-term structure, we would have the scaling
\be\label{RRother}
\frac{Q^2 b^{2n}}{L^{4n}} \sim  \frac{1}{x^2} \frac{1}{L^{2\tilde n}}\Rightarrow p_{\beta}=\frac{1}{2-(2\tilde n+K)/2n}\,.
\ee
Here we  have used the fact that the Ramond-Ramond terms come with a relative factor of $g_s^2$ compared to the $H_3^2$ and curvature terms.
We know that $\tilde n\ge 0$ (this is the usual statement that sources dilute at large radius in string theory), and  we are not aware of complete scenarios for perturbative moduli stabilization with $K<0$.
Thus, presently known constructions of this sort of flux stabilization do not populate the range $0<p_{\beta}<1/2$ (or equivalently $p_f>0$).
Instead, for sufficiently large positive $2\tilde n+K$ we have $p_{\beta}< 0$ here, leading to case (1).  This is what happened in the above example with $p_f= -1/3$ \cite{powers}.
On the other hand, as we have noted, the case $p_f>0$ immediately arises in other versions of monodromy inflation, such as the brane-induced case described above.

Another relation giving potential bounds on $p_f$ is
\be\label{ppttwo}
p=(1+p_f)p_{V}\,,
\ee
where $p_{V}$ is the power computed by incorporating the backreaction of the inflationary energy on the moduli potential, and the corresponding flattening, before taking into account a nontrivial power from the kinetic term (\ref{phib}).   In other words, $p_V$  is the power that incorporates flattening from the potential, and $p$  is the final power that also incorporates kinetic flattening.  If $p_V$ is a fixed positive number, then through (\ref{ppttwo}), observational limits on the primordial perturbations imply an upper limit on $p_f$.
However, strong potential-flattening effects are a logical possibility: $p_{V}$ could in principle be very small to compensate a large value of $p_f$.  That is, one could have strong potential flattening combined with strong kinetic steepening.  It would be interesting to analyze  more general axion monodromy  scenarios  in order to characterize this possibility.  In extreme limits --- such as large total dimensionality $D$ --- one can obtain large fiducial powers in the higher-dimensional Lagrangian, and it would be worthwhile to characterize the ensuing powers $p, p_f, p_\Lambda$ in those cases \cite{SC}.

Let us recap the pattern we have discerned from this exploration.  The `kinetic inversion' case (3b), $p_{\beta}>1/2$, which leads to a monomial with a negative power, is observationally unviable.
The marginal case $p_{\beta}=1/2$, which leads to an exponential potential, is also unviable.
The kinetic steepening case $0<p_{\beta}<1/2$ could lead to viable or unviable examples, depending on the power $p_{V}$ (\ref{ppttwo}) appearing in the potential expressed in terms of the basic variable $b$. In other words, viability depends on how steep the potential starts out, and how much further it gets steepened.  However, as just explained,  at present we have no examples of mechanisms that would produce examples with $0<p_{\beta}<1/2$.  Next, examples with $p_{\beta}<0$, corresponding to $p_f<0$, arise quite readily in  power-law stabilization scenarios,  e.g.~in the example of \S\ref{motivatingexample}.  Finally, examples with $p_f>0$ can arise from brane-induced monodromy.  It would be interesting to  understand whether the sign of $p_f$ could be used to distinguish the two  classes of constructions described above, or if instead our current exploration is too simplistic for such distinctions.

\subsubsection{Range of frequencies for power-law stabilization}

In this section we will provide simple estimates of the range of frequencies of interest in simple power law scenarios.
At sufficiently high values of $f/M_p$, oscillations become degenerate with parameters determining the overall amplitude and tilt of the power spectrum.  From this, one might find interesting effects related to the low-$\ell$ behavior of the power spectrum as studied previously \cite{oscillationsdata}; however, the statistical significance at this point is low.  For string compactifications involving single axions,
one typically finds $f\lesssim M_p$, with theoretically controlled examples satisfying $f/M_p\lesssim 10^{-1}$.    In the case of multi-axion mechanisms, such as \cite{multiax}, larger values of $f/M_p$ are possible.

The lower end of the range of $f/M_p$ is of particular interest.  One threshold that arises is the value of $f$ at which a single-field effective field theory description breaks down. Imposing that the frequency in time of oscillations be less than the unitarity scale $\sim 4\pi f$ leads to a lower bound on $f/M_p$ of order $10^{-4}$.  The full theory may continue to produce oscillatory features beyond this scale, but their functional form becomes more difficult to calculate.  A practical threshold arises from the resolution available in the data.

Here we will make a similar rough estimate of the lower end of the range of $f/M_p$ from the string-theoretic point of view.  We will start with the case of single-scale power law stabilization, with the axion $b$ derived from the Neveu-Schwarz two-form potential $B$ discussed above.  After making that estimate, we will generalize to scenarios in which the axion might arise from higher-rank gauge potentials.

For large-field axion monodromy inflation we have (cf.~eqn (1.38) of \cite{LiamDanielBook})
\be\label{Hvalues}
H=3\times 10^{-5} \sqrt{\frac{r}{0.1}}M_p\,\qquad\Rightarrow\qquad \frac{H}{M_p}\gtrsim 10^{-5}\,.
\ee
In string compactifications from $D$ to $4$ dimensions, the four-dimensional Planck mass is given by
\be\label{Mp}
M_p^2\sim \frac{(2\pi L\sqrt{\alpha'})^{D-4}}{g_s^2 \kappa^2}\,,
\ee
where $L$ is the size of the extra dimensions in string units, $1/\alpha'$ is the scale of the string tension, and $g_s^2\kappa^2$ is the higher-dimensional Newton's constant.  For example in $D=10$, to which we will specialize, $2\kappa^2=(2\pi)^7 {\alpha'}^4$.

If we go to the boundary of control, imposing only that $H\sqrt{\alpha'}\le 1/L$ we can write
\be\label{floose}
\frac{f}{M_p}\sim \frac{1}{L^2} \ge H^2 \alpha' \sim L^6 (H/M_p)^2 \times (1/g_s^2)\,.
\ee
Again we may go to the marginal case of $g_s\to 1$ to get a conservative\footnote{By conservative here we mean avoiding imposing potentially overly strong theoretical priors that are consequences of our computational limitations.}  bound.  Also taking $H\sim 10^{-5} M_p$,  (\ref{floose}) gives roughly $ L^8\lesssim 10^{10} \Rightarrow f\gtrsim 10^{-5/4} M_p$.

However, we find a larger window if we allow for the possibility of axions from higher-rank gauge potentials.  In that case, we have $f/M_p\sim  L^{-r_a}$, where $r_a$ is the rank of the potential field.  The most extreme case would be a potential of rank $D-4$.   In this case, for $D=10$ the inequality (\ref{floose}) gives $f/M_p\gtrsim 10^{-5}$.

Altogether, our top-down theoretical priors are not very informative about the lower end of the range of $f/M_p$.  As such, it is well motivated to scan down at least to the unitarity scale, based on EFT reasoning, or even more agnostically to scan down to the resolution available in the data.

\subsection{Nonperturbative stabilization}\label{sec:nonperturbative}

In some cases, the moduli whose adjustments during inflation cause drift in the period may be stabilized via a balance of nonperturbative and perturbative terms in the potential.   In this section, we will explore a few such scenarios.  They produce a more general functional form of the drift $f(\phi)$ than  that  obtained in (\ref{powersform}); as such, these scenarios require use of the broader template (\ref{Texp}), keeping more parameters in the Taylor expansion.

\subsubsection{Oscillations in NS5-brane axion monodromy}  \label{KKLTexample}

We will begin by examining a specific realization of axion monodromy in a nonperturbatively-stabilized compactification of type IIB  string theory \cite{MSW}.  In this scenario the dimensionless axion corresponding to the inflaton is given by
\begin{equation}
a \equiv \int_{\Sigma_c} C_2\,,
\end{equation} where $C_2$ is the Ramond-Ramond two-form potential, and $\Sigma_c$ is a homology class corresponding to a family of two-cycles.
Situating the family $\Sigma_c$  in a warped region and wrapping an NS5-brane and an anti-NS5-brane on  well-separated representatives of $\Sigma_c$ introduces a monodromy that weakly breaks the shift symmetry of $a$.  The resulting potential is asymptotically linear, $V = \mu^3\phi$, with $\phi = a f$, where $f$  is given by
\begin{equation}
\frac{f^2}{M_p^2} = \frac{g_s}{8\pi^2}\frac{c_{\alpha c c} v^{\alpha}}{{\cal V}}\,. \label{KKLTf}
\end{equation} Here $v^{\alpha}$, $\alpha=1,\ldots h^{1,1}_+$,  are the volumes of two-cycles in the compactification; ${\cal V}$  is the total six-volume; and $c_{\alpha c c}$  are triple intersection numbers.  See \cite{FMPWX}, whose conventions we adopt here, for more detailed background.

The structures required for NS5-brane monodromy are compatible with moduli stabilization in the KKLT \cite{KKLT} scenario, in which K\"ahler moduli are stabilized by  nonperturbative superpotential terms from Euclidean D3-branes and/or from gaugino condensation on D7-branes.
The decay constant is determined by a specific combination of two-cycle volumes $c_{\alpha c c} v^{\alpha}$, and by the overall volume ${\cal V}$.
The inflaton potential  depends on ${\cal V}$ via
\begin{equation}
\mu^3 = const.\times {\cal{V}}^{-2}\,.
\end{equation}
This is an example of the universal decompactification instability caused by positive energy sources in four dimensions.   As a result, ${\cal V}$ couples strongly to the changing inflationary energy, diminishing as inflation proceeds, and we can estimate the shift in $f$ by examining the shift in ${\cal V}$.

We will consider a toy model that is a variant of the  two-modulus example given in \cite{MSW}.   To capture the shift of ${\cal V}$, it suffices to work in a simplified model with a single K\"ahler modulus $T$.\footnote{The shift of the two-cycle whose volume is denoted $v_+$ in \cite{MSW} induces a subleading correction to the shift in $f$, for the parameter values we will consider.}
We take
\begin{equation}
W = W_0 + {\cal A} e^{-a T}\,, \qquad  K = - 3\,{\rm{log}}(T+\bar T)\,,   \label{KKLTdata}
\end{equation}  with $W_0= 3\times 10^{-2}$, $a=\frac{2\pi}{25}$, ${\cal A}=-1$, as in \cite{MSW}.
This theory has a supersymmetric $AdS_4$  minimum at $\sigma \equiv {\rm{Re}}(T) \approx 21$, and can be uplifted to a metastable de Sitter vacuum through the addition of supersymmetry-breaking vacuum energy, e.g.~from an anti-D3-brane.   Upon further including\footnote{A K\"ahler  moduli space of dimension $>1$  is required for this purpose, as explained in \cite{MSW}, but this does not preclude  making an accurate estimate for $h^{1,1}_+=1$, as we do here.} a  sufficiently warped NS5-brane/anti-NS5-brane pair, axion monodromy inflation can occur without causing decompactification.
However, for sufficiently large inflationary energy $\mu^3\phi = V_{crit}$, as will arise if the warp factor approaches unity,  the quasi-de Sitter minimum disappears and decompactification ensues.
The maximum drift of the frequency  of oscillations occurs in a model that is almost unstable to decompactification at the start of inflation: the  shift of ${\cal V}$ is then the largest possible change compatible with K\"ahler moduli stabilization.

The location $\sigma_\text{min}$ of the minimum, as a function of $\phi$, is accurately captured by a second-order Taylor expansion around a point $\phi_0$ in field space,
\begin{equation}
\sigma_{\star}(\phi) \approx \sigma_\text{min}(\phi_0) + \sigma^{\prime}|_{\phi_0}(\phi-\phi_0)+\frac{1}{2}\sigma^{\prime\prime}|_{\phi_0}(\phi-\phi_0)^2\,.
\end{equation}
If the volume ${\cal V}$ is taken to be almost unstable at the start of inflation, and $\phi_0$  is taken to be near the point where the pivot scale exits the horizon,
then for the model parameters quoted above, one finds $\sigma^{\prime}|_{\phi_{0}} \lesssim 0.5$ and $\sigma^{\prime\prime}|_{\phi_{0}} \lesssim 0.1$.
These are (approximate) upper limits: if the relative sizes of the moduli potential and the inflationary energy are such that ${\cal V}$ is robustly stabilized at the start of inflation, considerably smaller shifts are possible.
Next, expanding $f$ as in (\ref{Texp}),
\begin{equation}
f(\phi)/f_0 \approx 1 + \frac{1}{f_0}\frac{df}{d\phi}\Bigr|_{\phi_{0}} (\phi-\phi_0)+\frac{1}{2 f_0}\frac{d^2f}{d\phi^2}\Bigr|_{\phi_{0}}(\phi-\phi_0)^2\,,  \label{quadratictemplate}
\end{equation}  and using (\ref{KKLTf}), we find  that for the same assumptions made above, $\frac{1}{f_0}\frac{df}{d\phi}|_{\phi_{0}} \sim 10^{-2}$ and $\frac{1}{2 f_0}\frac{d^2f}{d\phi^2}|_{\phi_{0}} \sim 10^{-3}$. One can then check that for the parameters of this toy model, $\Delta f/f \lesssim 0.03$ over the range of inflation visible in the CMB.
Even though the fractional change  in the frequency is small, for sufficiently small $f$ this effect may be observable.

We conclude that in realizations of NS5-brane axion monodromy in KKLT compactifications with realistic parameter values, the shifts of the  nonperturbatively-stabilized K\"ahler moduli, while small, can in some cases lead to detectable changes in frequency over the observed CMB.   The drifting frequency in this model is directly captured by (\ref{quadratictemplate}).

\subsubsection{Logarithmic drift  from nonperturbative stabilization}

We now point out that in some classes of nonperturbatively-stabilized compactifications, there is a logarithmic correction to the drift in frequency.
Very schematically, in terms of one relevant modulus $L$, nonperturbative stabilization  can lead to a relation of the form
\be\label{NPbalance}
\mu_V^4 e^{-L^n}\sim \mu_a^4 \frac{a^{p_0}}{L^\gamma}\,,
\ee
with $\mu_a^4/(\mu_V^4 L^\gamma) \equiv \epsilon_a L^{-\gamma} \ll 1$ small so that the two terms can compete.\footnote{For example, in KKLT  constructions this small parameter  is the fine-tuned small value of the classical flux superpotential $W_0$.}
This translates into
\be\label{NPbalancelog}
L^n \sim -\log(\epsilon_a a^{p_0}/L^\gamma) \sim -\log(\epsilon_a a^{p_0})\,.
\ee
Combining (\ref{NPbalancelog}) with $f=M_p L^{-n_f}$,
we find that the sinusoidal dependence is of the form
\be\label{NPosc}
\cos\Biggl[ b_0\,\phi\,\Bigl(-{\rm{log}}(\epsilon_a)-p_0\,{\rm{log}}(\phi/\phi_0)\Bigr)^{n_f/n}\Biggr]\,,
\ee where $b_0$  is a dimensionful constant. Here we have used  the fact that at least in a simple situation with a single overall scale, we have $a^{p_0}/L^\gamma\sim(\phi/M_p)^{p_0}$.

In (\ref{NPosc}) the exponent $n_f/n$ is positive, and might be expected to range over a few discrete rational values, roughly of order unity.\footnote{In supercritical theories one could have large $n_f$ and small $n$ for some axions.}
The presence of the log factor inside the cosine changes the template considerably, leading to a qualitatively different $\ell$-dependence of the oscillations.

\subsubsection{A supergravity toy example}

To explore possible relations among the parameters appearing in the oscillatory templates, it will be instructive to consider additional examples.
We will therefore examine oscillations in a simple ${\cal N}=1$  supergravity theory inspired by those that arise in nonperturbatively-stabilized string compactifications.\footnote{Whether the moduli potential arises from perturbative or nonperturbative effects will turn out to be of little importance in this example, but the fact that the Lagrangian is supersymmetric will be central.}
In this example the moduli-dependence of the axion decay constant (setting the period of a modulation) will be related to the moduli-dependence of a dynamically-generated scale  (setting the amplitude of the modulation).

Consider an ${\cal N}=1$  supergravity theory with a single  chiral superfield $T=\tau+i\theta$  and K\"ahler potential $K$.
First we will discuss the case in which the modulations come from a superpotential term, and periodic terms in the K\"ahler potential can be neglected.
Modulations in $K$  will be treated below.  Taking $\theta$  to be the candidate axion-inflaton, we suppose that
\begin{equation}
K(T,\bar T) = K(\tau)\,.
\end{equation}
We take the modulation to arise from a superpotential coupling,
\begin{equation}
W= W_{0} + W_1 e^{-\alpha T}\,,
\end{equation}  where $W_{0}$, $W_{1}$ are  constants independent of the modulus $T$,  and $\alpha \in \mathbb{R}$.
To be concrete, suppose that in Planck units,
\begin{equation}
K_{T\bar T} = \tau^{-2\beta}\,,
\end{equation}  with $\beta \in \mathbb{R}$, up to an overall constant  that we will not retain.
For $\beta=1$  this supergravity theory captures key structures arising in flux compactifications of type IIB string theory on Calabi-Yau orientifolds, with $T$  corresponding to a K\"ahler modulus.
The canonically-normalized  field is (neglecting $\tau-\theta$ kinetic mixing)
\begin{equation}
\phi = \theta \,\tau^{-\beta}\,.
\end{equation}
If the nonperturbative term is subleading overall, i.e.~if $W_{1} e^{-\alpha T} \ll W_0$, then the leading modulated term is
\begin{equation}
V_{\rm{osc}} = W_0 W_1 e^{-\alpha \tau} {\rm{cos}}(\alpha\theta) \equiv  \Lambda^4(\tau)\, {\rm{cos}}(\alpha\tau^{\beta}\phi) \,,
\end{equation} up to a phase that we omit.
So in full we have
\begin{equation}
V_{\rm{osc}} = \Lambda^4\Bigl(\tau[\phi]\Bigr)\, {\rm{cos}}\Bigl(\alpha(\tau[\phi])^{\beta}\phi\Bigr) = \Lambda_0^4 \, e^{-\alpha (\tau[\phi]-\tau_0)} \,, {\rm{cos}}(\alpha(\tau[\phi]-\tau_0)^{\beta}\phi)\ ,
\end{equation} where we allow the saxion $\tau$  to depend on the axion vev, $\phi$.
Notice that the scale and amplitude of the modulations are linked, both depending on $\tau[\phi]$, because the amplitude determined by the real part of an instanton action,  while the period is determined by the imaginary part of the same action (and, at the same time, by the kinetic term).
Linearizing in small shifts of $\tau$,
\begin{equation}
\tau[\phi]=\tau_0 + \gamma \phi\,,
\end{equation} with $\gamma$ a dimensionful constant, we  find
\begin{equation}
V_{\rm{osc}} = \Lambda_0^4 \, e^{-\alpha \gamma \phi} \, {\rm{cos}}\Bigl(\alpha \gamma^{\beta} \phi^{1+\beta}\Bigr)\,.
\end{equation}
Comparing to  the template (\ref{powersform}), we have $p_{\Lambda}=1$, $p_f=\beta$.

Next, we turn to modulations arising from instanton corrections to $K$ (this is, for example, a leading source of modulations in the NS5-brane construction of axion monodromy inflation in \cite{MSW,FMPWX}.)
Taking as a concrete example (cf.~\cite{MSW})
\begin{equation}
K = - 3 \, {\rm{log}}\,(T+\bar{T}+e^{-T}+e^{-\bar{T}})\,,  \label{modulatedk}
\end{equation} and neglecting nonperturbative terms in the superpotential, one finds a structure of the schematic form
\begin{equation}
\frac{e^{-\tau}}{\tau}\, {\rm{cos}}\left(\frac{\phi}{f}-\frac{e^{-\tau}}{2\tau} {\rm{sin}}(\phi/f)\right)\,,
\end{equation}
or more generally
\begin{equation}
\epsilon\, {\rm{cos}}\left(\frac{\phi}{f}-\epsilon\, {\rm{sin}}(\phi/f)\right)\,.  \label{eq:cosofsin}
\end{equation}
We learn that deviations from a pure sinusoidal form can arise if the oscillations in $\Lambda$ and in $f$ are correlated, e.g.~if both originate from a single correction to the K\"ahler potential.
However,  significant deviations from the sinusoidal template occur only when the modulations are large ($\epsilon \sim 1$) and a number of our approximations are breaking down.

\subsubsection{Enhanced drift from nonperturbative potentials}

To understand whether the relative rigidity of moduli stabilization in the example of \S\ref{KKLTexample}, and the correspondingly small drift, are general consequences of nonperturbative stabilization, it is useful to examine the dependence on canonically-normalized fields.  From the K\"ahler potential (\ref{KKLTdata}), the canonically-normalized field $\varrho$ corresponding to ${\rm{Re}}(T)$ is
\begin{equation}
\varrho = \sqrt{\frac{3}{4}}\,{\rm{log}}(T+\bar T)+C_{\varrho}\,,
\end{equation} with $C_{\varrho}$ an integration constant.   Comparing to (\ref{KKLTf}), we see that $f$ has an exponential dependence on $\varrho$.
This finding is fairly general: the axion kinetic term typically has exponential dependence on the canonically-normalized fields that parameterize `universal' fluctuations,  such as the overall volume ${\cal V}$  and the string coupling $g_s$.   In supersymmetric compactifications, this is a straightforward consequence of the logarithmic K\"ahler potentials of these fluctuations, as in (\ref{KKLTdata}).

However, it is worth asking, still in the context of nonperturbative stabilization, whether $f$ could have power-law dependence on a  more general modulus $\psi$ that does not parameterize a universal fluctuation. As an example motivated by the Large Volume Scenario \cite{LARGE}, consider a relative K\"ahler modulus  in an example where the K\"ahler  potential for two K\"ahler  moduli $\tau_b$, $\tau_s$ takes the form
\begin{equation}
K=-2\, {\rm{log}}({\cal{V}}) = - 2\, {\rm{log}}(\tau_{b}^{3/2}-\tau_{s}^{3/2})\,.
\end{equation}
In the regime $\tau_b \gg \tau_s$, $\tau_s$ is to good approximation  proportional to a power of a canonical field: one can expand the  logarithmic K\"ahler  potential.

In nonperturbatively-stabilized compactifications, it is often the case that leading terms in the  scalar potential have steep dependence on $\tau_s$, so that we have schematically
\begin{equation}
f\sim \tau_s^{\alpha},\qquad V \sim e^{-\tau_s}\,,
\end{equation}
for $\alpha$ a constant.
It follows that the relative shift of $f$ can be large when $\tau_s \ll 1$,
\begin{equation}
\frac{\delta f}{f} \sim \frac{\partial_{\tau_s} f}{f} \frac{\partial_{\tau_s}V_{infl}}{\partial^2_{\tau_s}V_{mod}} \sim \frac{\alpha}{\tau_s} \frac{V_{infl}}{V_{mod}} \,, \label{largeshift}
\end{equation} where $V_{infl}$  and $V_{mod}$  denote the inflaton potential and the moduli potential, respectively.

We conclude that a shift in $\tau_s$  that is modest in Planck units can be a large shift relative to the initial vev, and give rise to a large relative change in $f$.   The dynamics set by the inflationary and moduli potentials can be thought of as determining the shift measured in Planck units,  while the functional form $f(\tau_s)$ (e.g., exponential  or power law) determines whether such a shift  leads to a  small or large multiplicative change in $f$.   It would be interesting to identify examples in which the large  frequency shifts (\ref{largeshift}) lead to distinctive signatures.

\section{The Amplitude of Oscillations}\label{sec:creationmyth}

In this section, we will comment on the amplitude $\Lambda^4(\phi)$ of oscillations in  the potential (\ref{genpot}).  There is a wide range of possibilities for contributions to the part of the potential periodic in $a(\phi)$.   These contributions are generally exponentially suppressed as a function of the internal length scales in the extra dimensions, so it is important to emphasize that the amplitude of the effect is quite possibly unobservably small.

\subsection{The amplitude and the scale of inflation}

In high scale inflation there are significant upper bounds on the sizes of the extra dimensions, leading to reasonable regimes of parameters for which  oscillations could be detectable.
We will now make some simple estimates, in the spirit of the  preceding discussion, to illustrate the relation between the amplitude of oscillations and the geometry.  As well as helping to frame our understanding of the motivation for the search for oscillatory features, these estimates may provide some utility in interpreting the implications of a null result for axion monodromy and related scenarios.

As reviewed above,  axions can arise from potential fields integrated over topologically nontrivial submanifolds (`cycles') $\Sigma_I$ in the extra dimensions.  The spectrum of branes wrapping these cycles is periodic under $a_I\to a_I+2\pi$.  The brane masses depend on $a_I$ as well as on the volumes $v_I$ of the cycles.  If a cycle size is relatively small, then as the axion rolls along its trajectory, new sectors of  wrapped branes become light.  This can lead to particle/defect production, or  to radiative corrections, which produce periodic modulations, reflected in the oscillatory term in $V(\phi)$ (\ref{genpot}).   Similarly, instanton effects generically contribute oscillatory terms to the potential.

Each of these effects is exponentially suppressed as a function of the size of the cycle contributing the axion.
A large oscillatory contribution can arise if the corresponding cycle is stabilized at a small value --- even a vanishing value is tractable theoretically under some circumstances.
For a small or vanishing cycle size, one finds a value of $f/M_p$  that is independent of $v_I$.  It is therefore well motivated to search for oscillations over the whole range of values of $f/M_p$ for which the data has sufficient resolution.

However, the signal may even be detectable for  relatively large cycles, given the constraints on internal dimensions required for large-field inflation.  Let us make some estimates for the case studied above in \S\ref{sec:powerlaw}, with one scale $L\sqrt{\alpha'} \equiv \sqrt{v\alpha'}$ in the internal geometry.  In this case, for example for an axion from the Neveu-Schwarz two-form potential field $B$, we have worldsheet instanton effects that
contribute
\be\label{Lambdawsinst}
\Lambda^4(\phi)\sim \Lambda_0^4 e^{-v}\,,
\ee
where for power law stabilization of moduli in a product structure, we can estimate the prefactor as
\be\label{prefactor}
\Lambda_0^4\sim \frac{g_s^4}{V_{tot}^2}\times \frac{1}{g_s^2}\times L^4 M_p^4 \sim {V}_{\cal R}\sim V_{mod}\,.
\ee
In this prefactor, the first factor is the conversion to Einstein frame, the second the standard $1/g_s^2$ scaling of a classical effect in spacetime, and the factor of $L^4$ represents the integral over the bosonic zero modes (transverse position) of the worldsheet instanton in the case of a simple product structure.  We should stress that the estimates we are doing here are meant to give a rough idea of the possible scale of the effect, rather than specific predictions of explicit models.
Here $V_{mod}$ refers to a typical term in the moduli-stabilizing potential, such as the contribution $V_{\cal R}$ obtained by dimensionally reducing the internal curvature.

In this type of scenario, we may also trade $v$ for $f/M_p$ via
\be\label{onescalesizes}
\frac{f}{M_p}\sim \frac{1}{(2\pi L)^2}\sim \frac{1}{v_a}\,.
\ee
This exhibits a simple relation $p_\Lambda = p_f$, but this relation is not general.

Next, we note from previous work \cite{oscillationsdata}\ that an analysis using current CMB data is sensitive to an oscillatory piece roughly of size $\Lambda^4(\phi)\sim 10^{-4} V_0(\phi)$.
Putting this together with the previous estimates, we have detectability for
\be\label{detectability}
e^{M_p/f}\le 10^{4} \frac{\Lambda_0^4}{V_{0}(\phi)}\,,
\ee
where $V_0$ is the scale of the inflationary potential energy.  Since $\Lambda_0^4$ can be of order the moduli-stabilizing potential barriers, it is at least as large as the inflationary potential $V_0(\phi)$.  Even in the marginal case, where they are equal, we see from (\ref{detectability}) that one requires $f/M_p\ge 1/9$, meaning that the cycle size $v$ is of order 10 in string units.  For high-scale inflation, this is a reasonable number.  And again, given a hierarchy between the inflationary and moduli-barrier scales, the amplitude would be detectable for even larger cycle size despite the exponential suppression.

Turning this around, a null result at this level would exclude an interesting regime of model parameters, given the bounds on the size of internal cycles arising from the high scale of inflation.  However, it would be difficult to falsify the scenario based on these constraints.  In any case, the estimates here provide ample motivation to pursue a search for drifting oscillations over a broad range of parameters.

\subsection{Creation myth: tunneling initial conditions and thresholds for $\Lambda(\phi)^4$}

In this section, we will make a slight detour to explain a relation between drifting oscillations and an interesting
scenario for the initial conditions (somewhat similar to those explored in \cite{unwinding}) in axion monodromy inflation.  An appealing possibility for the precursor to the $\approx$ 60 $e$-folds of phenomenological inflation is tunneling from some earlier metastable minimum in the effective potential.  This can be motivated by the structure of the string landscape, although despite interesting efforts there is not anything close to a well-understood framework for computing and interpreting physical observables in eternal inflation.

With  the oscillations studied here,  there are regimes of parameters and field values for which the oscillations produce local minima in the potential.  In order for phenomenological slow-roll inflation to proceed, we require that the potential be monotonic during the last 60 $e$-folds.
That is, we require $V_0'(\phi)$ to dominate at late times.
But if the derivative of the oscillatory term in the potential, $V_{osc}'(\phi)$, dominates at earlier times, i.e.~further out in $\phi$, tunneling out of a resulting metastable minimum could provide initial conditions for slow roll inflation.  Since the field range of $\phi$ is bounded by ultraviolet physics, extending only over the range $\phi<\phi_{UV}$ where the inflationary potential remains consistent with moduli stabilization \cite{MonodromyI, MSW, FMPWX, powers}, the condition that $V_{osc}'(\phi)$ dominates at large $\phi$ leads to a {\it lower} bound on the amplitude of oscillations.  This bound depends on the parameters $p, p_f, p_\Lambda$.

\subsubsection{Rigid case}

As an initial illustration, consider $p_\Lambda=p_f=0$ (no drift in amplitude or phase), and $V_0(\phi_{UV})\sim 10^2 V(\phi_0)$.

A convenient parameterization is
\beq
V'=p\, \mu^{4-p}\phi^{p-1}\left[1-b_p \sin\left(\frac{\phi}{f}+\gamma_0\right)\right]\,, ~~~~ b_p\equiv \frac{\Lambda^4}{p\, \mu^{4-p}\phi^{p-1}f}\,.
\eeq
Having a tunneling `creation myth' induced by oscillations requires that the secular term $\mu^{4-p}\phi^{p-1}$ dominates $V'$ for $\phi<\phi_0$ for slow-roll to work, while at some value $\phiCDL>\phi_0$ the oscillatory contribution to $V'$ must begin to dominate and create local minima to tunnel from. That is, we need
\beq
b_p> 1 ~~~~ {\rm for} ~~~~ \phi>\phiCDL ~~~~ {\rm and} ~~~~ b_p< 1 ~~~~{\rm for} ~~~~ \phi<\phi_0\,.
\eeq
Here $\phiCDL$ denotes the point of the (last) tunneling exit, which is in the region where $b_p$ crosses from $b_p>1$ to $b_p<1$ with decreasing $\phi$. That is, $b_p$ must grow monotonically with increasing $\phi$, increasing beyond unity around $\phi\sim\phiCDL>\phi_0$. From the form of $b_p$ this immediately requires $p<1$ for the rigid case we are discussing right now.

Furthermore, there is a finite range of control $\phiUV<\infty$ above which the energy density grows too large.
Hence we need to require $\phi_0<\phiCDL\leq\phiUV<\infty$ for a tunneling based creation myth to work. This fact that $\phiCDL\leq\phiUV<\infty$ immediately enforces a \emph{lower} bound on $\Lambda$.

We can analyze this more closely by rewriting the above conditions $b_p(\phiCDL)>1$ and $\phiCDL\leq\phiUV$ explicitly as a lower bound on $\Lambda$
\bea
\Lambda^4 > \Lambda_c^4\equiv p\,f\mu^{4-p}\phi_0^{p-1}\left(\frac{\phiUV}{\phi_0}\right)^{p-1}&=&p\,\frac{V_0}{\phi_0} \left(\frac{V_0}{\VUV}\right)^{\frac{1-p}{p}} f\nonumber\\
&&\\
&=& 3\times 10^{-10}\MP^4\; p^{3/2}\left(\frac{V_0}{\VUV}\right)^{\frac{1-p}{p}} \frac{f}{\MP}\,.\nonumber
\eea
Here we used $\phi_0\simeq 10\sqrt{p}\, \MP$ and the power spectrum normalization
\beq
{\cal P}_{S}=\frac{1}{24\pi^2}\frac{V_0}{\MP^4\epsilon_0}=\frac{1}{12\pi^2}\frac{V_0\phi_0^2}{\MP^6}\,\frac{1}{p^2}=2.2\times 10^{-9}\quad\Rightarrow ~~~~ V_0\simeq 3\times 10^{-9} p\,.
\eeq
Observationally, oscillations are bounded in amplitude, giving a constraint
\beq  \label{experimentalbound}
\Lambda^4 < \Lambda_{exp}^4\sim 10^{-13}\MP^4\,\sqrt{\frac{f}{\MP}}\,.
\eeq
We now require
\beq
\Lambda_c^4<\Lambda^4<\Lambda_{{\rm{exp}}}^4
\eeq
for a model to provide a creation myth consistent with experimental data.
This rearranges into
\beq
p^{3/2}\,\left(\frac{V_0}{\VUV}\right)^{\frac{1-p}{p}} < 3\times 10^{-4}\sqrt{\frac{\MP}{f}}\,,
\eeq
In turn, this provides an upper bound on $p$
which leads to a constraint on $r=4p/N_e$.
To get a sense of the numbers in this case, let us allow $f/\MP$ to range down to $10^{-4}$ and estimate from the high scale of inflation that $V_0/\VUV\gtrsim 0.01$.  This leads to an upper bound $r\lesssim0.04$.

\subsubsection{Relaxing rigidity}

More generally, however, the additional parameters $ p_f, $ and especially $p_\Lambda$, in our drifting templates relax this condition. For each value of these parameters that is consistent with a tunneling initial condition, there is a lower bound on the amplitude $\Lambda(\phi_0)^4$, but it will vary significantly over  the range of interest.  The power $p_\Lambda$ enters exponentially in (\ref{powersform}).  Over the range $\phi_0<\phi<\phi_{UV}$, this can lead to a large hierarchy between $\Lambda(\phi_{UV})^4$ and $\Lambda(\phi_0)^4$, allowing for tunneling initial conditions.

We start from the expression for $V'$ descending from \eqref{powersform}, which we write as
\beq\label{Vprime}
V' =p\, \mu^{4-p}\phi^{p-1}\left[1-\bpf \sin\left(\frac{\phi_0}{f_0}\left(\frac{\phi}{\phi_0}\right)^{1+p_f}+\gamma_0\right)-\bpl \cos\left(\frac{\phi_0}{f_0}\left(\frac{\phi}{\phi_0}\right)^{1+p_f}+\gamma_0\right)\right]\,,
\eeq
where
\bea\label{eq:bp1}
\bpf&=&\frac{\Lambda^4(\phi)}{p\,\mu^{4-p}\phi^{p-1}f_0} (1+p_f)\left(\frac{\phi}{\phi_0}\right)^{p_f}\,,\\
&&\nonumber\\
\label{eq:bp2}
\bpl&=&\frac{\Lambda^4(\phi)}{p\,\mu^{4-p}\phi^{p-1}f_0} p_\Lambda C_0 \frac{f_0}{\phi_0}\left(\frac{\phi}{\phi_0}\right)^{p_\Lambda-1}\,,\\
&&\nonumber\\
\Lambda^4(\phi)&=&\MP^4\,e^{-C_0\left(\frac{\phi}{\phi_0}\right)^{p_\Lambda}}\nonumber\,.
\eea
Already at this level, we see that $p_\Lambda >0$ strongly disfavors having a creation myth, as in this case $\bpf,\bpl \to 0$ exponentially quickly with increasing $\phi>\phi_0$.
So we look at the parameter ranges $p_f\neq 0$, $p_\Lambda\leq 0$.

If we work in a regime where $p_f\sim p_\Lambda$ without a strong hierarchy between them (which the power-law cases above seem to support), then $\bpl\ll \bpf$ due to the extra power of $\phi_0/\phi$ and the factor $f_0/\phi_0\ll 1$.

The parameter ranges $p_f\neq 0$, $p_\Lambda\leq 0$ (neglecting $\bpl$ as argued above) provide two possibilities.

\begin{itemize}
\item case a: $p_\Lambda=0$, i.e.~only phase drift. In this case the creation myth conditions $\bpf(\phiCDL)>1$ together with $\phiCDL\leq\phiUV$ boil down to a modification of the rigid case condition
\bea
\Lambda^4 >\Lambda_c^4&\equiv &\frac{p}{1+p_f} \frac{V_0}{\phi_0} \left(\frac{V_0}{\VUV}\right)^{\frac{1+p_f-p}{p}} f_0\nonumber\\
&&\\
& =&3\times 10^{-10}\MP^4 \;\frac{p^{3/2}}{1+p_f} \left(\frac{V_0}{\VUV}\right)^{\frac{1+p_f-p}{p}} \frac{f_0}{\MP}\,.\nonumber
\eea
Consequently, in comparing with the results from the rigid case,
$p_f< 0$ strengthens this upper bound, while $p_f > 0$ weakens it. Again, we can get a feeling for numbers by taking the same numerical estimates as before,
$f=10^{-4}\MP$ and $V_0/\VUV=0.01$.
Then we get $r\lesssim 0.04\, (1+p_f)$ up to terms logarithmic in $1+p_f$ with small prefactor.

\item case b: $p_\Lambda < 0$, i.e.~both amplitude and phase drift. In this case the exponential dependence in $\Lambda^4(\phi)$ quickly dominates once $|p_\Lambda|\geq {\cal O}(0.1)$.

As an example, the model discussed above with $p_f=p_\Lambda=-1/3$, $p=4/3$ has a hierarchy
\be\label{Lratio}
\frac{\Lambda(\phi_{UV})^4}{\Lambda(\phi_0)^4} \sim {\rm{exp}}\left(\frac{M_p}{f_0}\left[1-\left(\frac{\phi_0}{\phi_{UV}}\right)^{1/3}\right]\right)\,.
\ee
Since $f_0/M_p\sim 1/L^2$ can be small, e.g.~of order $10^{-2}-10^{-1}$, this can easily exceed the ratio
\be\label{Vzeroratio}
\frac{V_0'(\phi_{UV})}{V_0'(\phi_{0})} \sim \frac{\phi_0^{1/3}}{\phi_{UV}^{1/3}}\,,
\ee
leading to a monotonic behavior during inflation and a tunneling initial condition.  Similar results hold for a significant range of these parameters.

The drift in $\Lambda$ described by $p_\Lambda$ has a limited effect on oscillations in the CMB, because of the small range of $\phi$ during the observable window.
However, we have just seen that this drift can have significant impact on tunneling initial condition.

\end{itemize}

\section{Symmetry Structure of Axion Monodromy}  \label{symmetry}

Now that we have explored the pattern of drifting oscillations, exploiting the backreaction of the inflationary potential on moduli degrees of freedom, it is worthwhile to summarize the symmetry structure of axion monodromy inflation.  The comments in this section can be found in the existing literature, but we collect them here for completeness, and because of the close connection between the discrete shift symmetry, its soft breaking, and the drifting oscillatory features in the resulting power spectrum.

There are several ways symmetries come into the physics.
First, there are the underlying gauge symmetries respected
by the generalized fluxes $\tilde F$ appearing in the Stueckelberg-like terms (\ref{Stuckelberg}).
As always, a gauge symmetry is a redundancy of description.  Nonetheless, the gauge symmetry
constrains the form of the action for the fluxes and their couplings to the gauge potential fields $A$ that descend to the axions $a_I\sim \int_{\Sigma_I}A$ via integration over cycles $\Sigma_I$.\footnote{As noted above, there are various string-dual fields, including examples from brane motion modes as in \cite{MonodromyI}\ or the T-dual `windup toy' picture of \cite{MSW,FMPWX}, which can behave similarly; another interesting class of examples can be found in \cite{unwinding}.}

In the presence of generic fluxes and/or branes, there is direct dependence on the axions in the four-dimensional effective potential.  A shift of $a(\phi)$ by $2\pi$ can be undone by a shift of flux quanta, giving an equivalent configuration.  But in a given sector of flux quanta --- quantum numbers which are locally stable, decaying via nonperturbative effects --- the potential is unwound to a kinematically unbounded field range.

If the system  is displaced from the minimum of the resulting axion potential, other scalar fields,  including moduli corresponding to the volume of the extra dimensions and to the string coupling $g_s$,  feel forces as a result.  These tadpoles are not periodic under the shift $a\to a+2\pi$.  The sourced fields adjust in an energetically favorable way, either by flattening or destabilizing the potential, as explained  for the case of power-law stabilization in \S\ref{sec:powerlaw}.  Their couplings in the axion kinetic term are also crucial, leading to kinetic steepening or flattening depending on the example.  In appropriate cases leading to viable phenomenology, the moduli remain metastable, with the adjustment leading to a flattened potential $V_0(\phi)\sim \mu^{4-p}\phi^p$.

Given that, there is a sector of particles and defects --- those arising from branes wrapping the cycle $\Sigma_a$ --- that respects a weakly broken discrete shift symmetry.   Much like the spacefilling branes that can contribute to the potential as in \cite{MSW,FMPWX}, branes of higher codimension also have tensions  that depend on the axion $a$,  and that undergo monodromy when $a \to a+2\pi$.  However, the full spectrum of such branes comes back to itself under $a\to a+2\pi$, up to small corrections inherited from the (generically flattened) potential $V_0(\phi)$.  Similarly, Euclidean brane configurations, which produce instanton effects, are periodic.  These sectors of the theory are important in the present work, as they generate the oscillatory features.

Finally, at low energies there is an effective approximate discrete shift symmetry.  This symmetry protects against substantial radiative corrections to the potential, as in standard large-field inflationary models such as \cite{Andreichaotic}.  This structure descends from gauge symmetries in the underlying string theory, and because of the sub-Planckian underlying periodicities, it is consistent with the  expectation that exact continuous global symmetries  should be absent in quantum gravity.

\section{Data Analysis}  \label{analysis}
We now turn to a search for the models we have discussed in the {\it{Planck}} nominal mission data. Our analysis is preliminary and should largely be thought of as a proof of principle, establishing that what we propose is numerically feasible.

The analysis divides into two parts: the computation of the primordial power spectrum, and a search for this power spectrum in the data. As we have argued in \S\ref{requiredprecision}, an analytic derivation of the primordial power spectrum will involve extending the derivation given in~\cite{FMPWX} to higher orders in the slow-roll expansion. This is beyond the scope of the current work, but there are at least three routes to avoid this tedious task. The first is a numerical computation of the primordial power spectrum, starting from the inflaton potential, at every point in parameter space. We will not pursue this option because it is very time consuming.  The second approach is to start from an analytic template based on the derivation in~\cite{FMPWX},  making use of a careful comparison of the template with a full numerical calculation for the interesting range of parameters to ensure that the two can be brought into agreement after small adjustments of existing parameters in the template. The third approach is a purely phenomenological template that captures the relevant physics. In this case the hard work  is to relate the parameters of the phenomenological template to those in the microscopic theory. We will present examples of both the second and third approaches, and give a dictionary between the parameters in the two approaches, in \S\ref{sec:temp}.

Once an accurate analytic template has been found, we can search for it in the data. The dimension of the parameter space to be searched is relatively high. There are the parameters of $\Lambda$CDM,
additional
parameters in the primordial power spectrum, and also nuisance parameters for foregrounds, relative calibration between frequencies, and beams. As a function of the nuisance parameters and the parameters of $\Lambda$CDM, the likelihood is typically smooth and not too far from Gaussian, and several techniques exist that allow for an efficient analysis. However, the likelihood typically depends very strongly on certain parameters of the oscillatory power spectrum, such as the frequency, and can have hundreds of local maxima, providing a numerical challenge. In particular, the standard Metropolis-Hastings algorithm fails and alternatives are needed. Early searches fixed the cosmological parameters and evaluated the likelihood function on a very fine grid in the frequency, in the remaining power spectrum parameters, and in the $\Lambda$CDM parameters that are degenerate with these. The advantage of this approach is
that
the entire parameter space is searched, and the grid spacing can be chosen fine enough so that no features are missed. The disadvantage is that searching the entire parameter space is inefficient because most regions are uninteresting. As a consequence it is not feasible to vary the cosmological and nuisance parameters. In models with linear or logarithmic oscillations, or the oscillations predicted by the linear axion monodromy model \cite{MSW,FMPWX}, the frequency of the oscillations is the only challenging parameter. So an alternative that allows varying all cosmological parameters is to use a Metropolis-Hastings sampler on many narrow slices in frequency. In
\S\ref{analysisresults}, we present the results of a search based on nested sampling as implemented by MultiNest~\cite{multinest}, which is designed to sample functions with many maxima. We demonstrate that this method would recover a signal using simulations based on an oscillatory power spectrum in
\S\ref{sec:sim}. To understand the improvement in the fit
that
we should expect in the absence of a signal, we also show simulations based on a smooth $\Lambda$CDM power spectrum.

\subsection{Templates and parameter ranges}\label{sec:temp}

Throughout this work, we focus on single-field axion monodromy. We will now specialize further to the inflaton potential (\ref{powersform}) motivated by power-law moduli stabilization. Using the same approximations as in \cite{FMPWX}, the power spectrum is
\begin{equation}\label{eq:Ppf}
\Delta_\mathcal{R}^2(k)=\Delta_\mathcal{R}^2\left(\frac{k}{k_\star}\right)^{n_s-1}\left(1+\delta n_s e^{-C_0\left(\frac{\phi_k}{\phi_0}\right)^{p_\Lambda}}\cos\left[\frac{\phi_0}{f}\left(\frac{\phi_k}{\phi_0}\right)^{p_f+1}+\Delta\varphi\right]\right)\,,
\end{equation}
where the amplitude $\delta n_s$ and the phase $\Delta\varphi$ are calculable using the techniques in~\cite{FMPWX}; $\phi_0$ is some fiducial value for the scalar field to be specified later; and $\phi_k$ is the value of the scalar field at the time when the mode with comoving momentum $k$ exits the horizon. At leading order in the slow-roll expansion we have
\begin{equation}
\frac{\phi_k}{M_p}=\sqrt{2p(N_0-\ln(k/k_\star)) }\,.
\end{equation}
where $N_0=N_\star+\left(\phi_{end}/M_p\right)^2/2p$, and $\phi_{end}$  is the value of the scalar field at the end of inflation.
In the spirit of our earlier discussions, we will for now neglect the drift in amplitude, setting $C_0=0$. Comparison with a numerical evaluation of the power spectrum shows that the  analytic template (\ref{eq:Ppf}) gives an excellent approximation to the exact power spectrum, after a small adjustment of the frequency. In other words, an accurate template for the power spectrum is
\begin{equation}\label{eq:temp1}
\Delta_\mathcal{R}^2(k)=\Delta_\mathcal{R}^2\left(\frac{k}{k_\star}\right)^{n_s-1}\left(1+\delta n_s \cos\left[\frac{\phi_0}{\tilde{f}}\left(\frac{\phi_k}{\phi_0}\right)^{p_f+1}+\Delta\varphi\right]\right)\,,
\end{equation}
with $\tilde{f}$ higher than the underlying axion decay constant by up to a few percent.  This is the first template for which we will present results in \S\ref{analysisresults}. The relation between the amplitude $\delta n_s$ and the parameters in the potential is well approximated by
\begin{equation}\label{eq:alpha}
\delta n_s=3 b\left(\frac{2\pi}{\alpha}\right)^{1/2}\qquad\text{with}\qquad\alpha=(1+p_f)\frac{\phi_0}{2f N_0}\left(\frac{\sqrt{2pN_0}}{\phi_0}\right)^{1+p_f}\,,
\end{equation}
and monotonicity parameter $b\equiv\tilde b_p^{(1)}(\phi_0)$ given by~\eqref{eq:bp1}.

\begin{center}
\begin{figure}[thb]\label{fig:pknumvsan}
\includegraphics[trim= 0.8cm 0cm 0cm 0cm,width=6.5in]{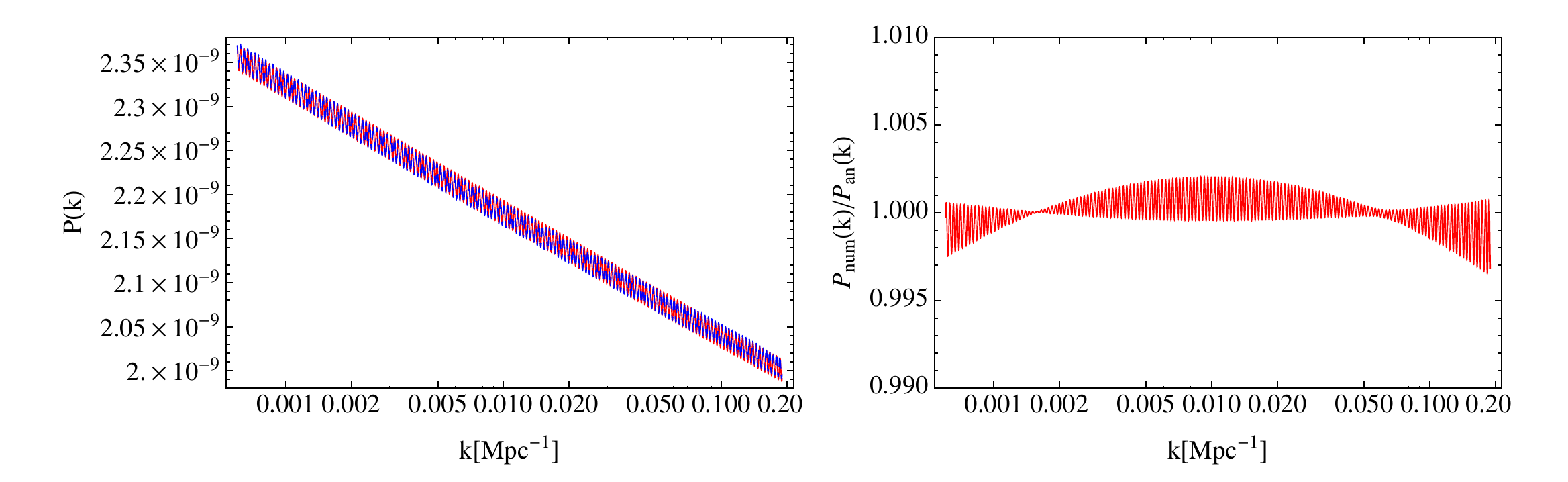}
\includegraphics[trim= 0.8cm 0cm 0cm 0cm,width=6.5in]{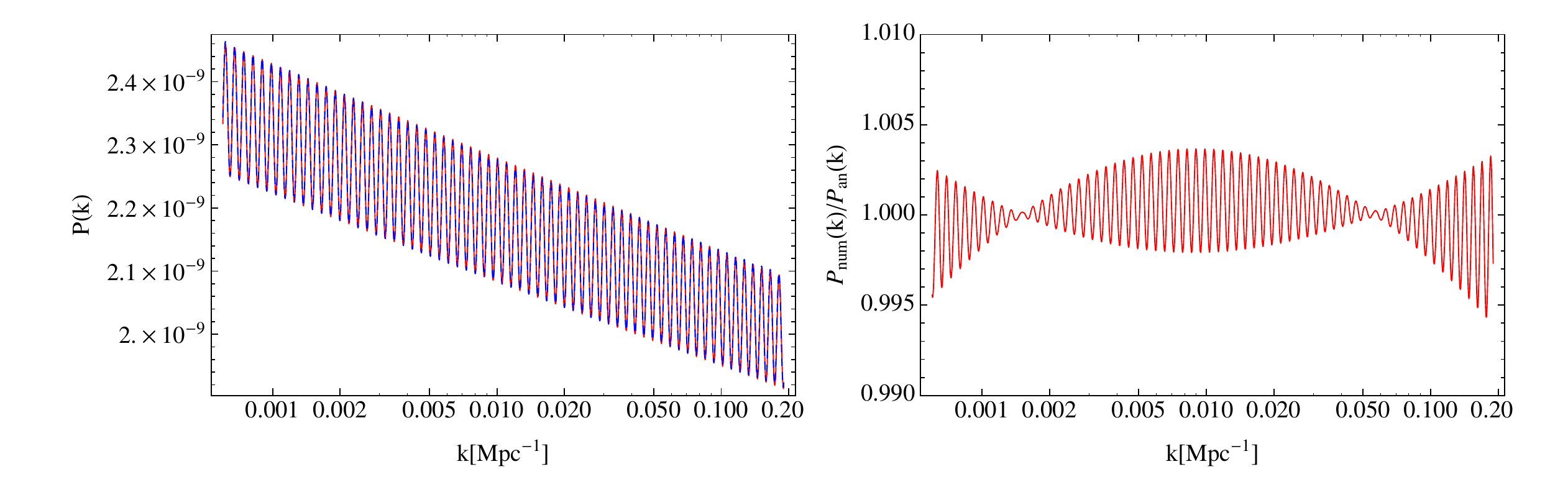}
\caption{Comparison of a numerical power spectrum and an analytic template after a small shift in $f$ in the analytic template to maximize the overlap.  Top row: $C_0=0$, $p_f=-\frac13$, $f=4\times 10^{-4}\,M_p$, and $b=0.01$.   Bottom row: $C_0=0$, $p_f=-\frac13$, $f=10^{-3}\,M_p$, and $b=0.05$. The top row is best viewed after zooming in.}
\end{figure}
\end{center}

Figure~\ref{fig:pknumvsan} shows the comparison between the numerical and analytic power spectra for two representative choices of parameters. For both we set $p=4/3$ and $p_f=-1/3$. For the first example, we set $f=4\times 10^{-4}\,M_p$ and $b=0.01$. In this case we find $\tilde{f}=4.16\times 10^{-4} \, M_p$. For the second example, we set $f=10^{-3}\,M_p$ and $b=0.05$ and find $\tilde{f}=1.03\times 10^{-3}\,M_p$. Since $f$ and $\tilde{f}$ are so close, we will omit the tilde in what follows.

It is instructive to expand the argument of the trigonometric function in $\ln(k/k_\star)$. One finds
\begin{multline}\label{eq:cosexp}
\cos\left[\frac{\phi_0}{f}\left(\frac{\phi_k}{\phi_0}\right)^{p_f+1}+\Delta\varphi\right]=\cos\Biggl[\alpha\left(\ln(k/k_\star)+\frac{1-p_f}{4N_0}\ln^2(k/k_\star)\right.\\\left.+\frac{(1-p_f)(3-p_f)}{2^2 3! N_0^2}\ln^3(k/k_\star)+\frac{(1-p_f)(3-p_f)(5-p_f)}{2^3 4! N_0^3}\ln^4(k/k_\star)+\dots\right)+\Delta\tilde\varphi\Biggr]\,.
\end{multline}

Even for the highest frequencies of interest, the term proportional to the fourth power of the logarithm does not significantly modify the phase over the observed seven e-folds, and can be neglected. However, we should in general keep the terms up to and including the third power of the logarithm to accurately model the power spectrum.  This is consistent with our findings in \S\ref{requiredprecision}, where we showed that we must retain terms quadratic in $\phi_k-\phi_0$ in the Taylor expansion of the axion decay function.

We see that the potential~\eqref{powersform} with $p_f$ of order unity leads to an argument of the trigonometric function that is a rapidly converging series in $\ln (k/k_\star)$. This suggests a natural phenomenological template of the form
\begin{equation}\label{eq:lnn}
\Delta_\mathcal{R}^2(k)=\Delta_\mathcal{R}^2\left(\frac{k}{k_\star}\right)^{n_s-1}\left(1+\delta n_s \cos\left[\Delta\varphi+\alpha\left(\ln(k/k_\star)+\sum\limits_{n=1}^2\frac{c_n}{N_\star^n}\ln^{n+1}(k/k_\star)\right)\right]\right)\,.
\end{equation}
This is the second template for which we will present results in \S\ref{analysisresults}. For a microphysical model  yielding a primordial power spectrum that is well described by (\ref{eq:lnn}), the $c_n$ will be related to the parameters of the model. These relations can be used to convert bounds on $c_n$ into bounds on microphysical parameters.

Modulations in the inflaton potential also lead to oscillatory contributions in the power spectrum of tensor perturbations. However, for $f\ll M_p$, the amplitude of these oscillations is suppressed compared to the amplitude of the oscillations in the scalar power spectrum,
by a factor $(f/M_p)^2\alpha\ll 1$~\cite{FMPWX,Flauger:2010ja}. So for practical purposes we can approximate the spectrum of tensor perturbations by the spectrum obtained in the absence of oscillations
\begin{equation}\label{eq:pkt}
\Delta_h^2(k)\approx\Delta_h^2\left(\frac{k}{k_\star}\right)^{n_t}\qquad\text{with}\qquad \Delta_h^2=r\Delta_\mathcal{R}^2\qquad\text{and}\qquad n_t=-r/8\,.
\end{equation}

To perform a search, we must specify priors for the various parameters in these templates. The parameters in our first template are $n_s$ (or $p$), the amplitude $\delta n_s$, $p_f$, $f$, and the phase $\Delta\varphi$. Concrete models with $p\in \{3,2,4/3,1,2/3\}$ have been explored in the literature, but more general powers are plausibly realizable.
As can be seen from~\eqref{eq:cosexp}, there is a degeneracy between $p$ and the axion decay constant,  so that changes in $p$ can be compensated for by changes in $f$. So $p$ is only directly constrained by the spectral index and the tensor-to-scalar ratio. With current data, the variation in $n_s$ and $r$ induced by changes in $p$ only leads to improvements of the fit of a few.
Thus, in an exploratory search for oscillatory patterns in the power spectrum, $p$ can be fixed; we will set $p=4/3$.
We will assume instantaneous reheating. For $p=4/3$ this corresponds to $N_\star\approx57.5$ for $k_\star=0.05\,{\rm Mpc}^{-1}$. We set $\phi_0=12.38\,M_p$ and find $\phi_\text{end}=0.59\,M_p$.
The best-fit amplitude in the WMAP9 data was large, $\delta n_s\approx 0.5$. We should choose a prior that includes this best-fit point; we will use $0<\delta n_s<0.7$. For  this initial analysis, we consider $-3/4 < p_f < 1/2$, but because we do not know of sharp theoretical bounds on the parameters $p$, $p_f$, we plan to perform scans over a wider range in the future.\footnote{Ongoing work to analyze the powers arising in axion monodromy inflation more systematically may yield more interesting bounds on, or relations among, the parameters.}  For the axion decay constant we scan over $10^{-4} M_p < f < 10^{-1}M_p$, and finally we take $0<\Delta\varphi<2\pi$.  For the second template, we assume that the parameters $c_n$ are of order unity and use $-2<c_n<2$, $1<\alpha<1000$, and $0<\Delta\varphi<2\pi$. It should be kept in mind that the single-field effective field theory becomes strongly coupled when $\alpha\gtrsim 200$. So a detection of a signal above this value would necessitate a better understanding of the microscopic model.

\subsection{Consistency tests with simulated data}\label{sec:sim}
Before we present the results of our search in the {\em Planck} nominal mission data, we will test our methods on simulated data.

The introduction of additional parameters into a model is expected to improve the fit. Parameters that enter the model linearly,
such as $\delta n_s$ in our case, are expected to improve the fit by $\Delta\chi^2=1$ per parameter. Parameters such as the axion decay constant are not of this type, and the expected improvement is harder to predict.
Roughly speaking, for amplitudes like $\delta n_s$ we only have one trial, and increasing the amplitude too far will lead to a bad fit. For parameters like $f$, however, an increase will not necessarily eventually lead to a bad fit. Some frequency may fit well,
and then as we increase the frequency the fit may become worse, but it may eventually become better again. In other words, we have several independent trials and should include a trials
factor for the look-elsewhere effect. While progress can be made analytically toward understanding the expected improvements for such parameters, we will resort to simulations based on a featureless power spectrum to assess the improvements we should expect for our templates in the absence of a signal. Our simulations include correlations between multipoles introduced by the sky masks used for the measurement of the spectra in the CAMspec likelihood code. In addition, the simulations include foregrounds consistent with the foreground model used in the CAMspec likelihood code presented in~\cite{Ade:2013kta}. They do not include instrumental effects, but we assume that instrumental effects are well enough controlled to be unimportant for the statistical question we are interested in. We have performed searches in five simulations for the template~\eqref{eq:temp1}, varying only the power spectrum parameters. The largest improvements found in the five runs range from $\Delta\chi^2\approx14$ to $\Delta\chi^2\approx19$. If these were drawn from a Gaussian distribution, we would be led to expect an improvement of $\Delta\chi^2=16.5$ with a standard deviation of $\sim3.5$. We know from the first reference in~\cite{oscillationsdata} that the distribution is not quite Gaussian, and more substantial simulations are required in the event of a potential detection to assess the significance. However, these numbers do provide a useful guide,
and improvements of this magnitude should not be considered significant.

We now demonstrate that our methods allow recovering a signal from simulated data based on an oscillatory power spectrum. As before, the simulations include effects of sky masks and foregrounds, but do not include any instrumental effects.
\begin{figure}[h!]
\begin{center}
\includegraphics[width=3.1in]{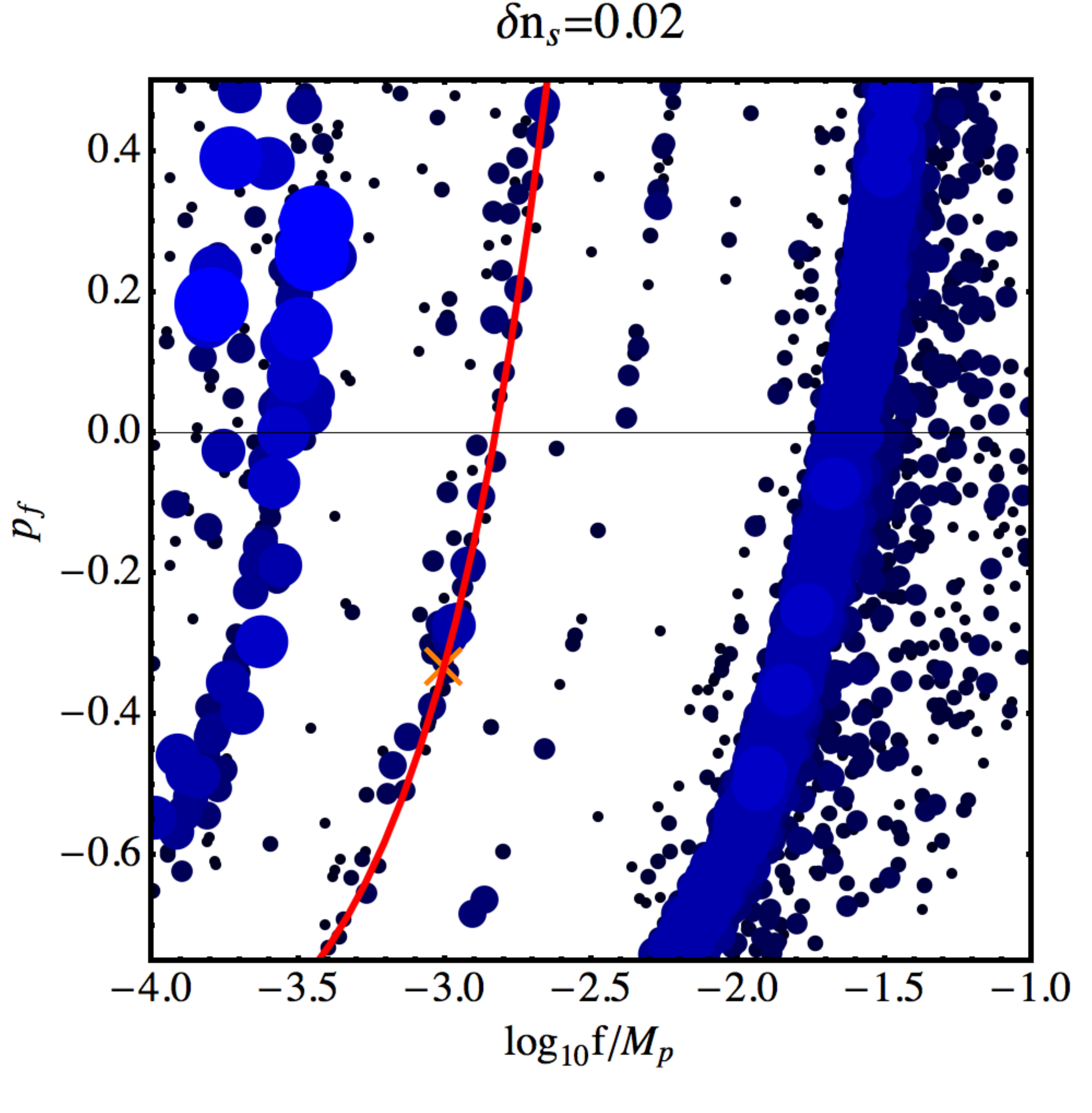}
\includegraphics[width=3.1in]{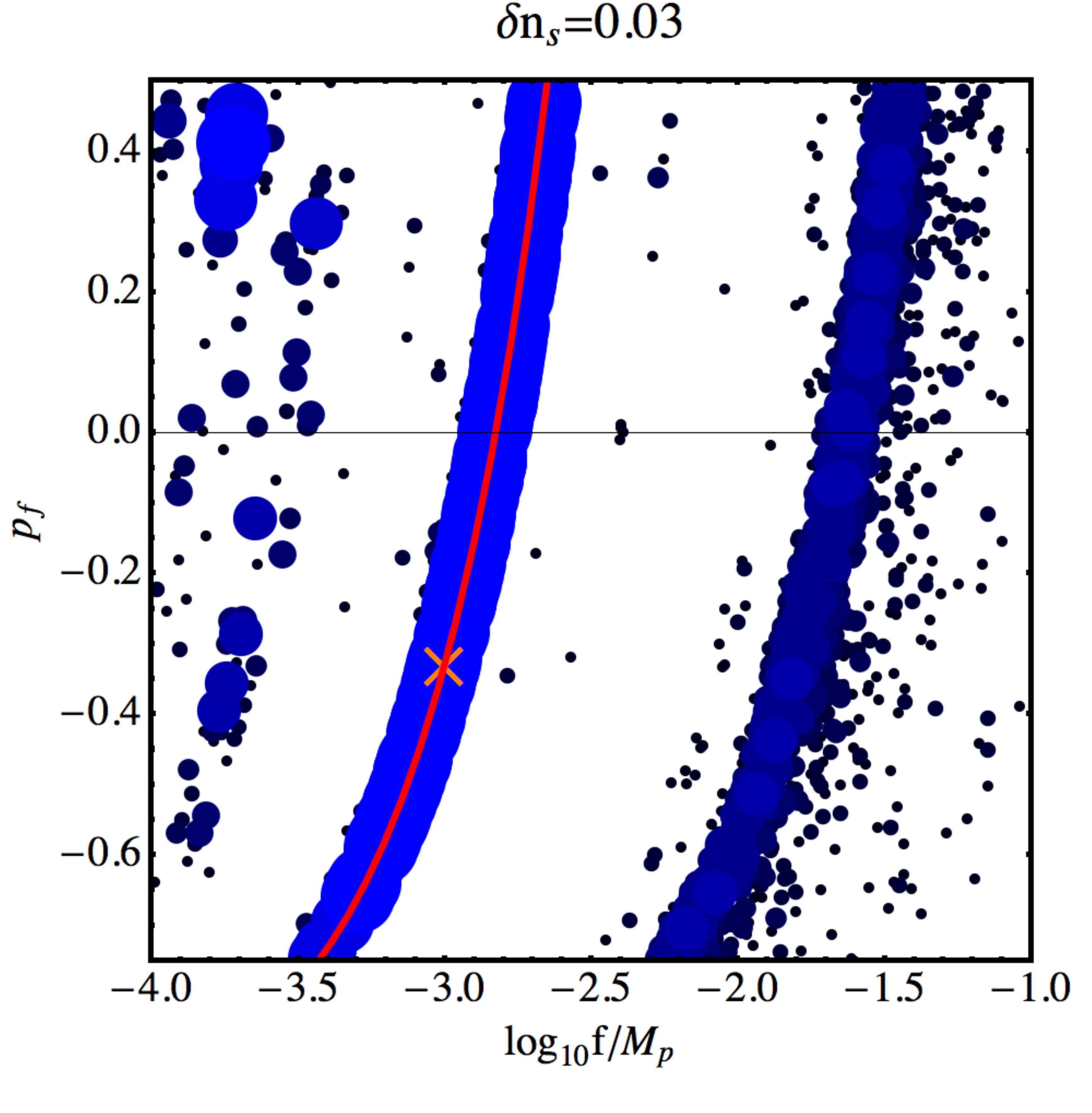}\\
\includegraphics[width=3.1in]{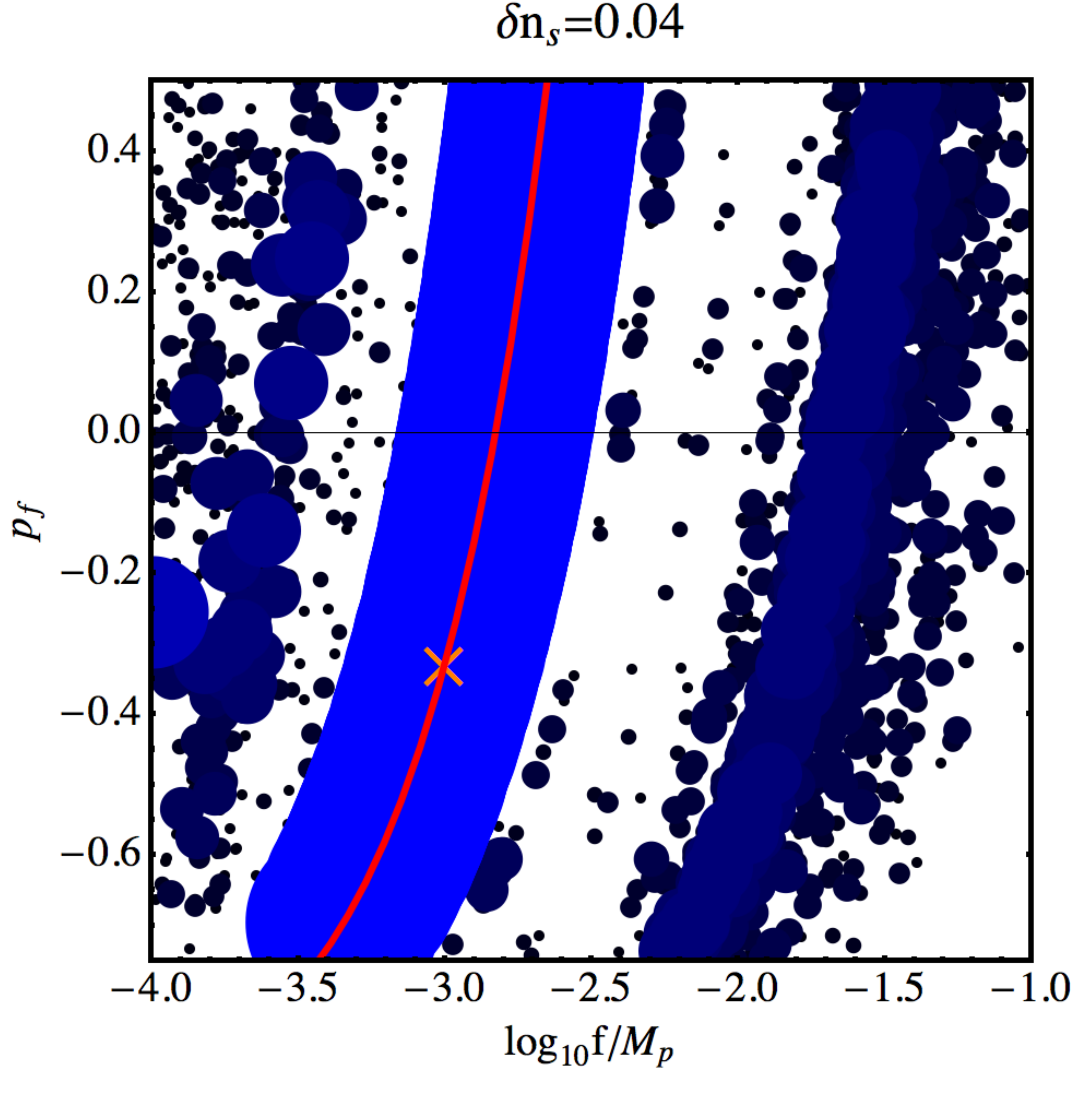}
\includegraphics[width=3.1in]{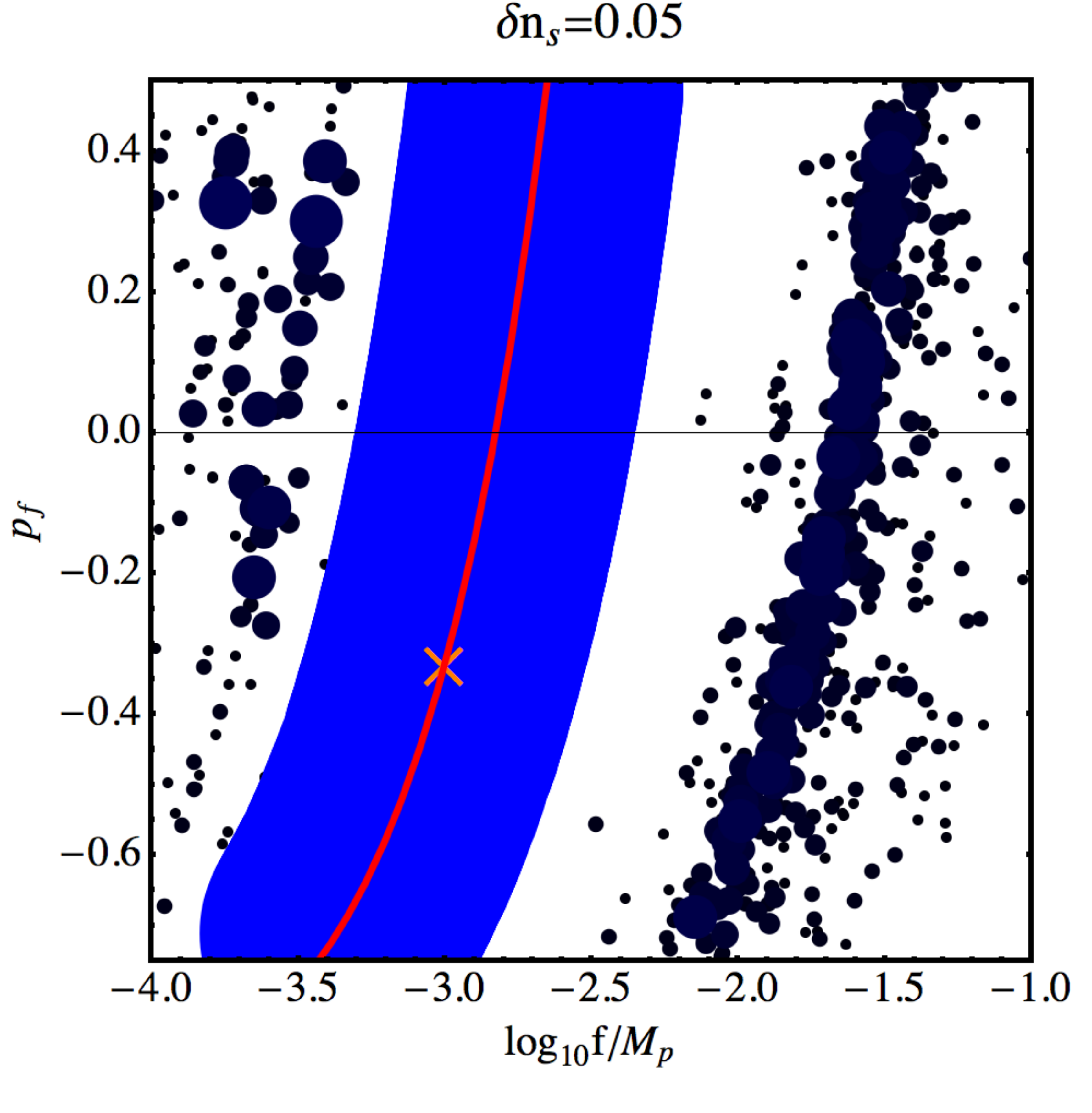}
\caption{Results for a search for oscillations in simulated data. The simulations are based on the template~\eqref{eq:temp1} with $p_f=-1/3$, $f=10^{-3}M_p$,
and amplitudes $\delta n_s$ ranging from $\delta n_s=0.02$ to $\delta n_s=0.05$,
as indicated by the labels. Points included in the plot lead to an improvement over the best-fit with $\delta n_s=0$ of $\Delta\chi^2\geq4$. The
sizes of the dots and their color indicate the amount of improvement. The orange cross indicates the input values of $f$ and $p_f$,
and the red line represents the corresponding value of $\alpha=\omega/H$ as defined in equation~\eqref{eq:alpha}. As the amplitude $\delta n_s$ increases the significance of the improvements
grows.
\label{fig:dnssim}}
\end{center}
\end{figure}
We have performed simulations for a signal of the form of template~\eqref{eq:temp1} with $p_f=-1/3$ and amplitudes ranging from $\delta n_s=0.02$ to $\delta n_s=0.2$ for three representative frequencies,
$f=2\times 10^{-4}\,M_p$, $f=4\times 10^{-4}M_p$, and $f=10^{-3}\,M_p$. We find that the signal is recovered in all simulations that lead to an improvement larger than that expected in the absence of a signal, but with
an amplitude $\delta n_s$ that is slightly biased towards larger values. For our preliminary search this is of no concern, but would have to be addressed in the event of a detection. The results are shown in
Figure~\ref{fig:dnssim} for the simulations with amplitudes $\delta n_s=0.02$, $\delta n_s=0.03$, $\delta n_s=0.04$, and $\delta n_s=0.05$ for axion decay constant
$f=10^{-3}\,M_p$ and $p_f=-1/3$. The four simulations are based on the same random seed,
so the only difference is the amplitude of the signal. The size and color of the circles represents the improvement. Larger circles represent larger values of $\Delta\chi^2$. For $\delta n_s=0.02$ the improvement corresponding to the input parameters is smaller than for other parameter choices that fit random scatter in the data. As the amplitude is increased to $\delta n_s=0.05$ the improvement grows to $\Delta\chi^2\approx 25$ and becomes significantly larger than the improvement expected in the absence of a signal. As expected from our discussion of overlaps, $p_f$ is not well constrained and $\Delta\chi^2$ is approximately constant along the line of constant $\alpha=\omega/H$ as defined in equation~\eqref{eq:alpha}.

\subsection{Preliminary results}  \label{analysisresults}

We now turn to the results of a preliminary search for oscillatory contributions to the power spectrum with drifting periods in the {\it{Planck}} nominal mission data using the data set referred to as {\it{Planck}}+WP~\cite{Ade:2013kta}.
In previous searches for logarithmic oscillations or oscillations motivated
by the linear axion monodromy model, with negligible drift of the axion decay constant, in the {\it{Planck}} nominal mission data using the public likelihood code, the improvement in fit was exactly as expected statistically (see figure 6 of the first reference in \cite{oscillationsdata}).  That is, the fit improved by $\Delta\chi^2\approx10$ over the best-fit featureless power spectrum, just as expected based on simulations that include the look-elsewhere effect.  In this work, we add parameters characterizing the drift in frequency, and would like to understand if they improve the fit in a significant way.

In summary, we find that the fit improves by $\Delta\chi^2\approx 18$ for both templates~\eqref{eq:temp1} and~\eqref{eq:cosexp}. Based on the simulations discussed in section~\S\ref{sec:sim}, this is comparable to the improvement expected in the absence of a signal. However, it should be kept in mind that we have studied a limited range of parameters and that a more thorough search is needed. For now, we present these preliminary results as part of our proof of principle for the feasibility of this data analysis rather than as a comprehensive search for models with drifting frequency.

\begin{figure}[thb]
\begin{center}
\includegraphics[width=3.1in]{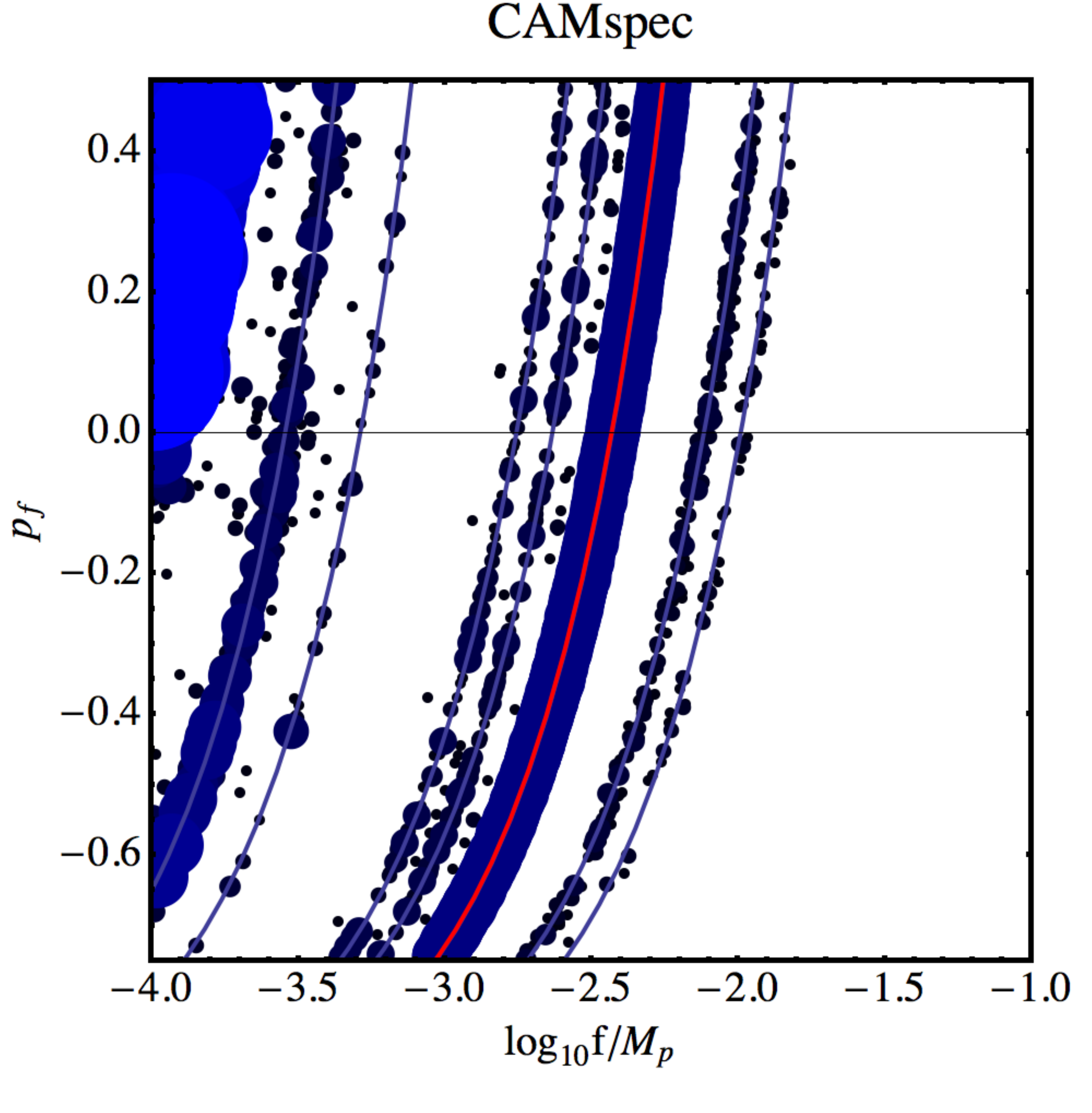}
\includegraphics[width=3.1in]{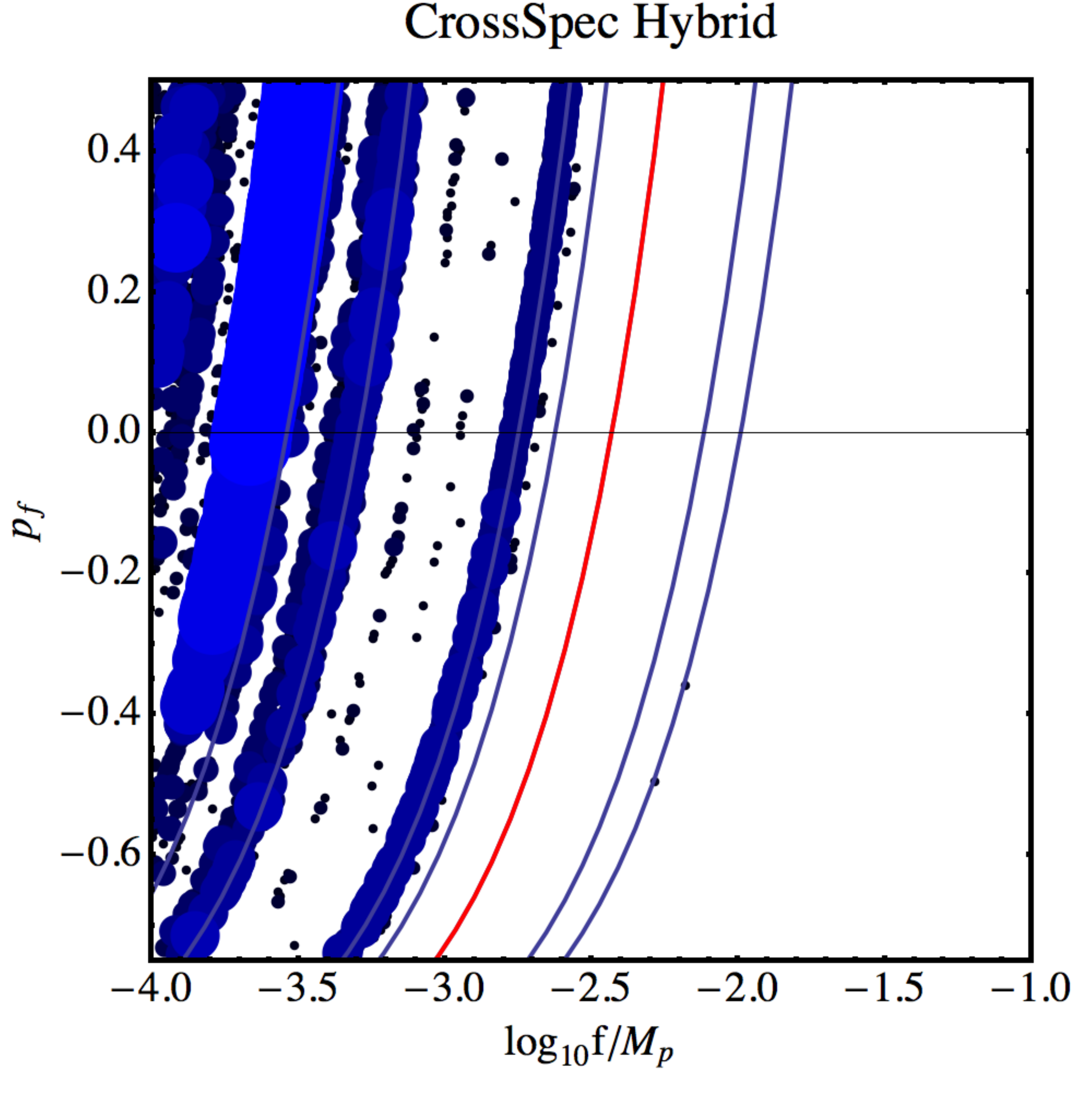}
\caption{Points that lead to an improvement over $\Lambda$CDM of $\Delta\chi^2\geq4$ for the first template,
\eqref{eq:temp1}. The left panel shows the results for the public {\it Planck} likelihood~\cite{Ade:2013kta},
and
the right panel shows the results for the likelihood discussed in~\cite{Spergel:2013rxa}. The
sizes of the dots and their color indicate the amount of improvement. Larger blue points represent larger values of $\Delta\chi^2$. The solid lines indicate constant $\alpha=\omega/H$. The red solid line represents $\omega/H=28.8$, the best-fit frequency found by the {\it{Planck}} collaboration~\cite{PlanckInf}. A significant contribution to the improvement derives from the region around $\ell=1800$ in the 217 GHz data. This range of multipoles of the 217 GHz data is known to be affected by interference between the 4K cooler and the read-out electronics~\cite{Spergel:2013rxa,Ade:2013kta}. \label{fig:temp1imp}}
\end{center}
\end{figure}

Let us now describe the analysis in detail. For our first template~\eqref{eq:temp1} we have performed three different types of runs, varying only the power spectrum parameters, varying the power spectrum and $\Lambda$CDM parameters, and varying the power spectrum and nuisance parameters in the CAMspec likelihood. If only the power spectrum parameters are varied, we find an improvement of $\Delta\chi^2\approx 18$. At the best-fit point, the axion decay constant is $f\approx 10^{-4}\,M_p$, $p_f=0.2$,
and the amplitude is $\delta n_s\approx 0.27$.
As expected, the improvement compared to $\Lambda$CDM is better when the $\Lambda$CDM or nuisance parameters are varied, but only slightly. Points that lead to an improvement of $\Delta\chi^2\geq4$ are shown in the left panel of Figure~\ref{fig:temp1imp}. We show the results for the run in which only power spectrum parameters are varied, as varying the cosmology or foregrounds tends to smear out the frequency.
The red solid line along which the fit is improved by $\Delta\chi^2\approx 10$ corresponds to a frequency $\omega/H=28.8$. This frequency was the best-fit identified by the {\it{Planck}} collaboration~\cite{PlanckInf}. As was pointed out in the first reference of~\cite{oscillationsdata}, some of the improvement can be attributed to a range of multipoles around $\ell=800$ and is seen in all frequencies. However, a significant contribution to the improvement originates from the 217 GHz data in the region around $\ell=1800$. This range of multipoles of the 217 GHz data is known to be affected by interference between the 4K cooler and the read-out electronics~\cite{Spergel:2013rxa,Ade:2013kta}. This disfavors a cosmological interpretation, but the agreement between analyses provides a consistency check for our search. The fact that the improvement is nearly constant along the line of constant $\omega/H$ shows that the data is not sensitive to the $\ln^2(k/k_\star)$ term for this relatively low frequency. At smaller $f$ or larger $\omega/H$ the improvement is no longer uniform along a line of constant $\omega/H$, so that the data prefers a certain amount of drift.

In addition, we have performed a search based on the likelihood described in~\cite{Spergel:2013rxa}.
This is based on survey cross-spectra and as a consequence is essentially insensitive to the cooler line but may be subject to other systematics such as changes over the course of six months. In this case, we find an improvement of $\Delta\chi^2\approx 14$ relative to the unmodulated power spectrum. The fraction of sky is slightly different between the two analyses and we have not performed simulations for our search
tailored to this likelihood, but we expect improvements comparable to those seen in the simulations for the CAMspec likelihood. The best-fit frequency in this case is $f=2.3\times 10^{-4} M_p$ with $p_f=0.05$ and an amplitude of $\delta n_s=0.15$. The results are shown in the right panel of Figure~\ref{fig:temp1imp}. The solid lines shown in the figure indicate the values of $\omega/H$ that led to improvements in the search based on the CAMspec likelihood. As expected, the improvement at $\omega/H\approx 28.8$ is absent, but there are some frequencies that lead to improvements in both
likelihoods, such as $\omega/H\approx60$ and $\omega/H\approx210$. For the linear axion monodromy model, the latter corresponds to an axion decay constant of $f=4.37\times 10^{-4} M_p$ and is the frequency that
led to the improvement of $\Delta\chi^2\approx 20$ in the WMAP9 data. The improvement seen in both likelihoods is $\Delta\chi^2\lesssim 10$.
\begin{figure}[b!]
\begin{center}
\includegraphics[width=3.0in]{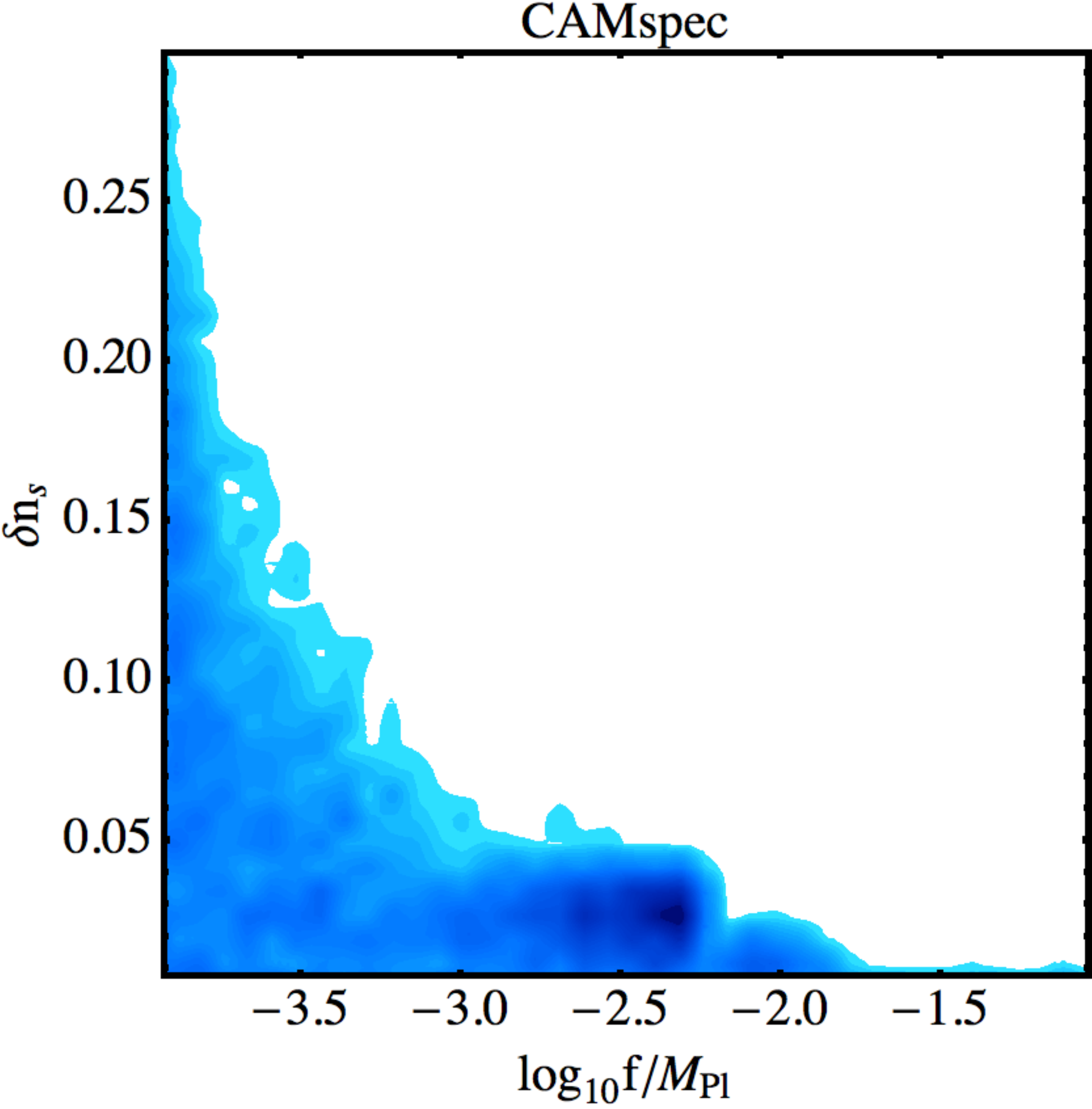}
\includegraphics[width=3.0in]{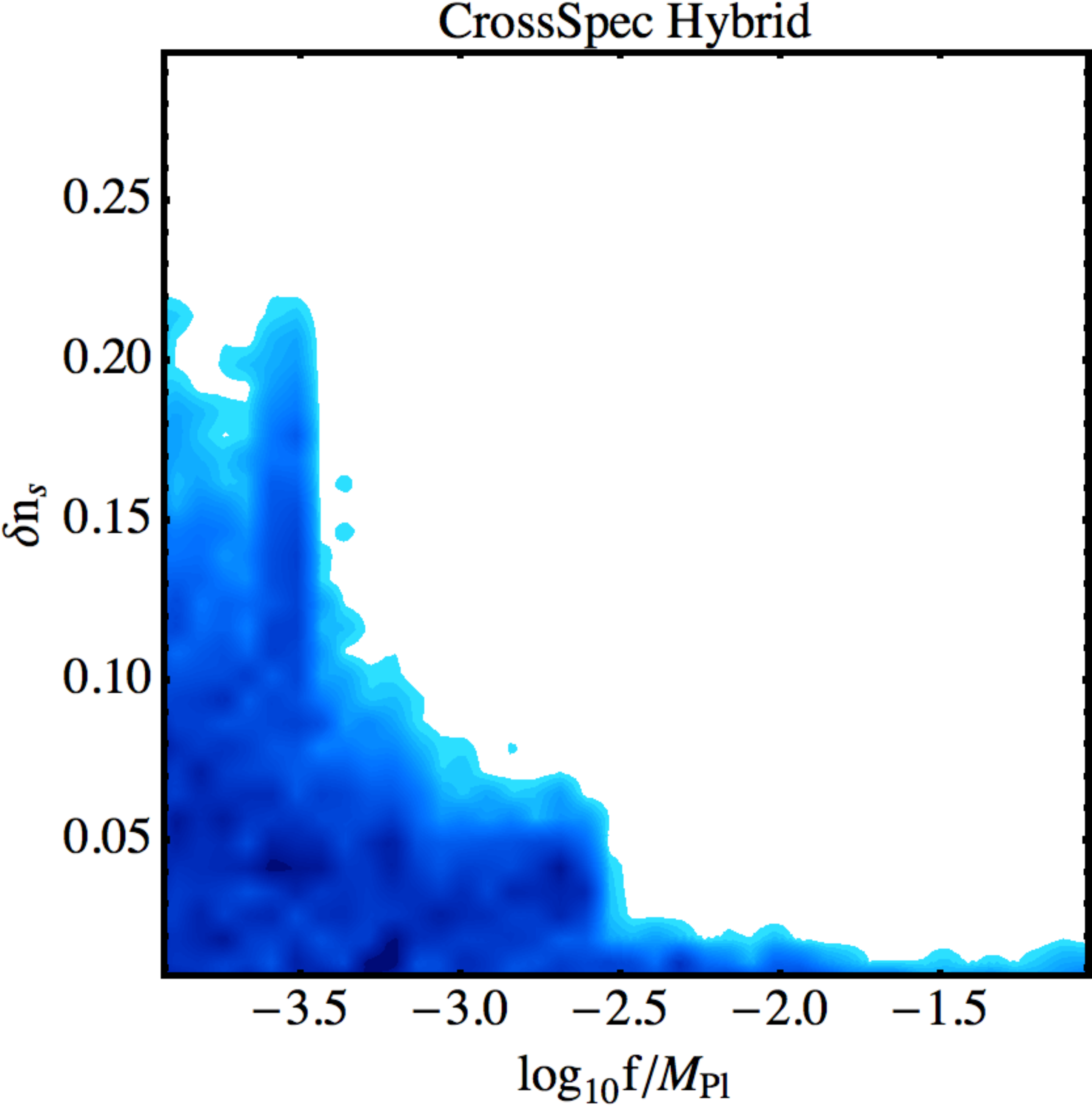}
\caption{Marginalized posterior in the $\delta n_s - f $ plane for the template~\eqref{eq:temp1},
for the run in which only the power spectrum parameters are varied. The left panel shows the results for the public {\it Planck} likelihood,
and the right panel shows
the results for the likelihood described in~\cite{Spergel:2013rxa}.  \label{fig:dnsvsf}}
\end{center}
\end{figure}

We can also use our searches to derive constraints on the amplitude of oscillations allowed by the data. The results are shown in Figure~\ref{fig:dnsvsf} for both likelihoods. When interpreting them, it should
be kept in mind both that
we restricted
to
$-3/4<p_f<1/2$ in our search, and that the limits are approximate given the limited number of simulations we have performed.

\begin{figure}[t!]
\begin{center}
\includegraphics[width=3.0in]{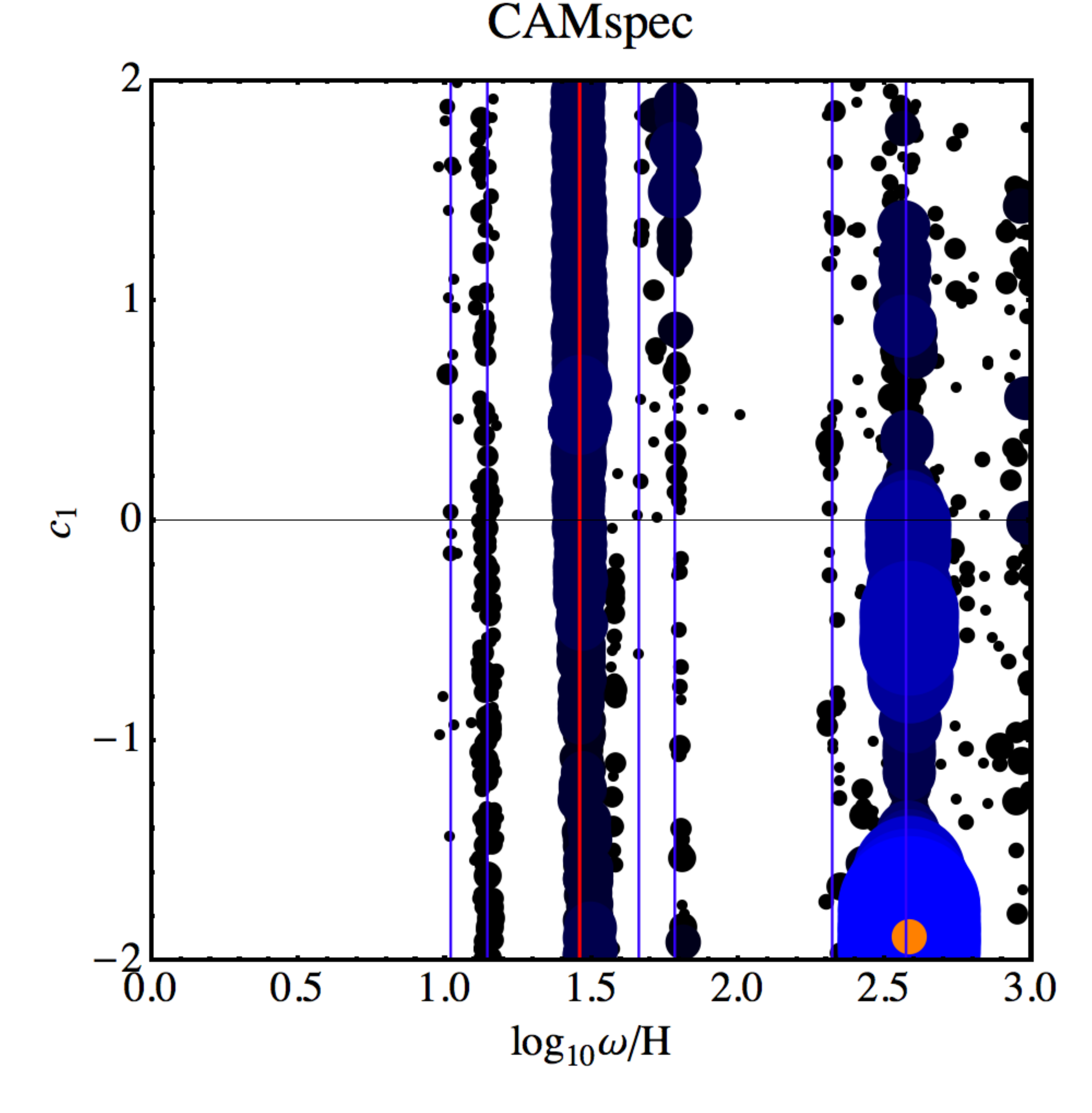}
\includegraphics[width=3.0in]{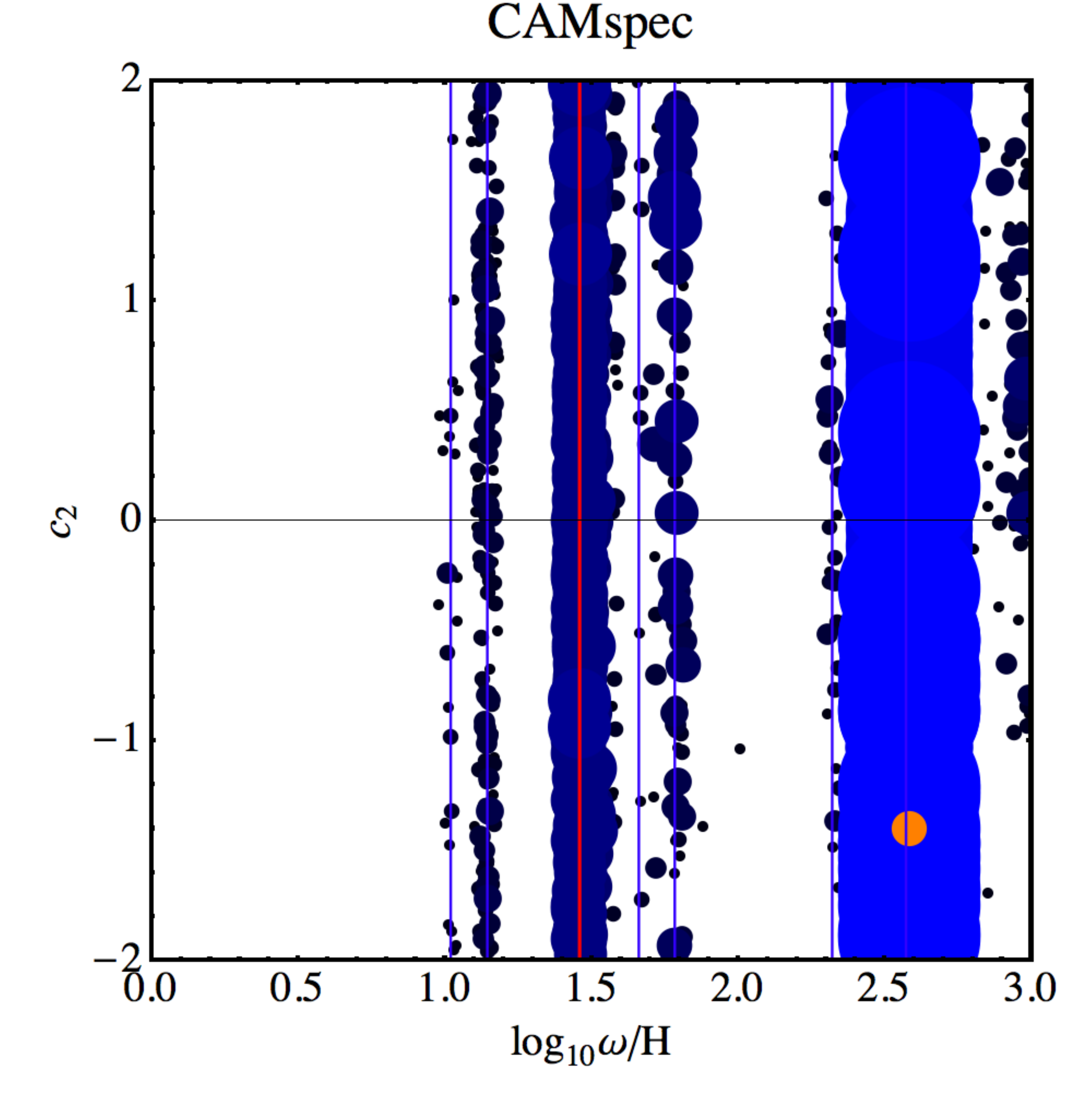}\\
\includegraphics[width=3.0in]{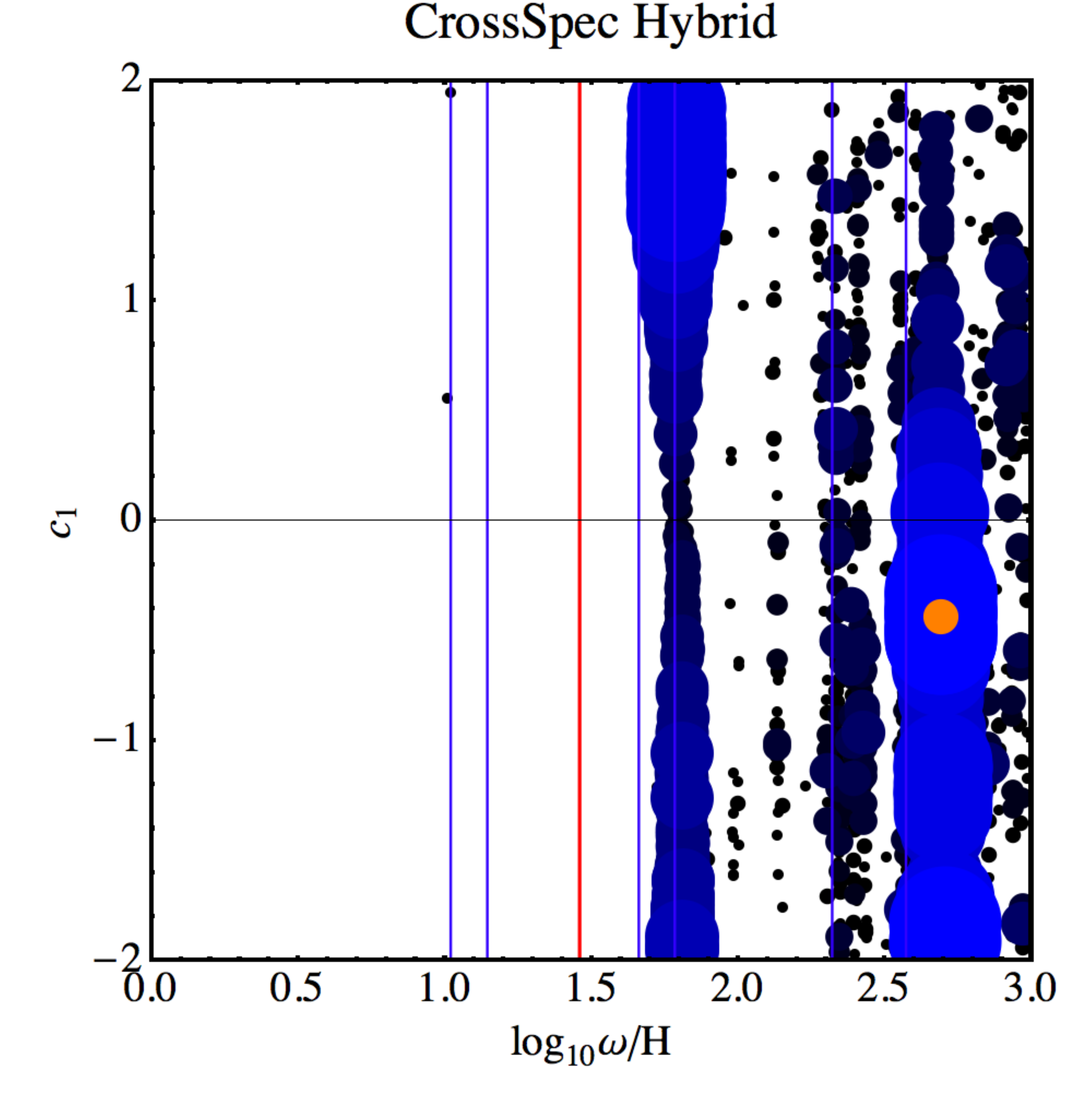}
\includegraphics[width=3.0in]{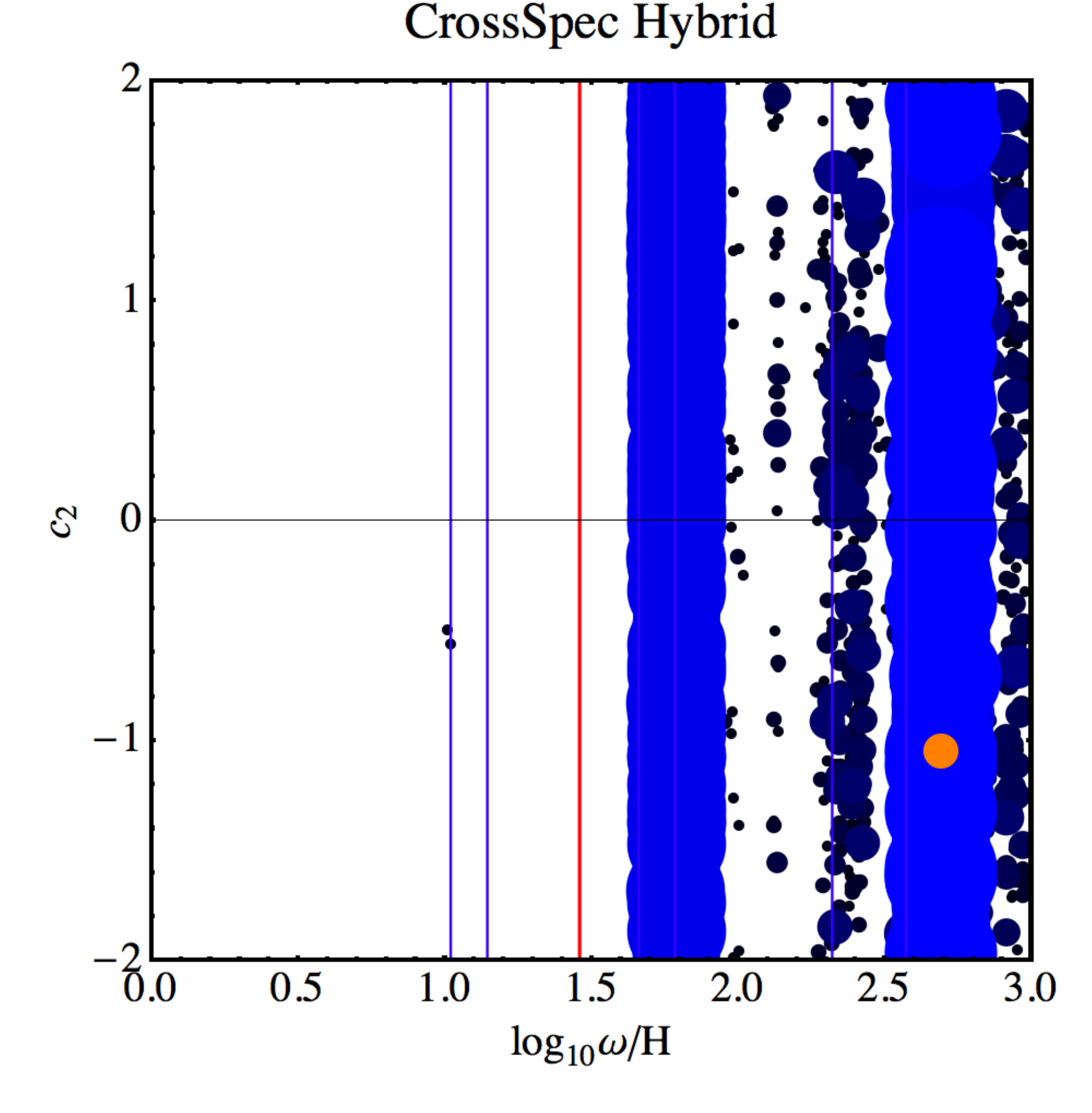}
\caption{Results for the second template, \eqref{eq:lnn}.
As before, points included in the plot lead to an improvement over $\Lambda$CDM of $\Delta\chi^2\geq4$,
and the sizes of the dots and their color indicate the amount of improvement. The orange dot indicates the best-fit point. The top panels shows the results for the public {\it Planck} likelihood,
and the bottom panels show
the results for the likelihood described in~\cite{Spergel:2013rxa}. While especially for
higher frequency there is a dependence on $c_1$, indicating that a drift in frequency leads to a better fit, the dependence on $c_2$ is weak.  \label{fig:logtemp}}
\end{center}
\end{figure}
The results for our second template~\eqref{eq:lnn}
are shown in Figure~\ref{fig:logtemp} for both likelihoods.
As before, no improvement is observed for $\omega/H=28.8$ for the likelihood described in~\cite{Spergel:2013rxa}, and improvements at $\omega/H\approx60$ and $\omega/H\approx210$ are seen in both likelihoods, but with $\Delta\chi^2\lesssim 10$. While the plots show a dependence on $c_1$ for the frequencies that lead to the most notable improvements to the fit, there is no dependence on the parameter $c_2$ for $\omega/H<1000$.

The improvement in the fit caused by instrumental effects such as the cooler line highlights that the search for oscillatory features in the power spectrum is subtle. Further evidence in this direction comes from the apparent discrepancy between the improvement seen in the {\it{Planck}} nominal mission data at $\omega/H\approx 210$ ($\Delta\chi^2\approx 10$) and the WMAP9 data ($\Delta\chi^2\approx 20$) in searches
for oscillations with constant frequency. Since the two experiments have very different resolution, a natural guess might be that one will recover the larger improvement in the {\it{Planck}} nominal mission data once the multipoles included in the analysis are restricted to the range in which both experiments are cosmic variance limited.
While the best-fit parameters in this case approach each other, the improvement in the analysis based on the CAMspec likelihood remains $\Delta\chi^2\approx 10$. A preliminary analysis \cite{RFCosmo} based on WMAP9 W-band data as well as {\it{Planck}} nominal mission 100 GHz survey spectra performed with the same masks and approximations to the covariance matrix led to an improvement $\Delta\chi^2\approx 10$ for the masks used in the {\it{Planck}} analysis in both datasets. As the sky fraction is increased, the improvement increases and approaches $\Delta\chi^2\approx 20$ for both the W-band and 100 GHz data. This indicates that a more thorough analysis is needed. 
Finally,
the range of parameters searched here was limited and a careful search should be carried out for a wider range of parameters and including the drift in the amplitude.

\section{Conclusions}  \label{theconclusion}

We have shown that the symmetry structure of axion monodromy inflation allows oscillatory terms in the scalar potential whose amplitude and frequency can drift over the course of inflation, as a consequence of moduli vevs that shift in response to the diminishing inflationary energy.   Incorporating the drifts in frequency can be critical in searching for oscillatory signatures in the scalar power spectrum.   We presented templates that encode the drifts that occur in a broad range of axion monodromy scenarios in string theory and in effective field theory.   The parameter ranges we propose for these templates include the values found in models such as \cite{MSW, FMPWX, powers}\ but extend well beyond this regime, including ranges in which the drift is strong enough to preclude a useful Taylor series expansion of the axion decay function,
but weak enough for radiative stability.
The range of frequencies and drift parameters consistent with an effective field theory description
includes regions where the drift parameters are essential in comparison to cosmological data, although for a wide range of parameters previous analyses without drift parameters suffice.  It would be interesting to pursue the theory of oscillations at higher frequencies, beyond the scale at which effective field theory breaks down, and to characterize the full range of parameters realized in string-theoretic axion monodromy.

Using the templates proposed here, we performed a preliminary search for drifting-frequency oscillations in the {\it{Planck}} nominal mission data, in order to demonstrate that such an analysis is numerically feasible.  A complete analysis that explores the full parameter space is an important problem for the future.

\section*{Acknowledgements}

We are very grateful to Hiranya Peiris for extensive discussions and correspondence, and to Richard Easther for early discussions of the role of the drift in $f$.
We would also like to thank Thomas Bachlechner,  Dick Bond, Fran\c{c}ois Bouchet, Xi Dong, Olivier Dor\'e, George Efstathiou, Daniel Green, Guy Gur-Ari, Ren\'ee Hlo{\v z}ek, Enrico Pajer, Leonardo Senatore, David Spergel, Gonzalo Torroba, and Herman Verlinde for useful discussions.
The work of L.M.~was supported by NSF grant PHY-0757868 and by a Simons Fellowship.
The work of E.S.~was supported  in part by the National Science Foundation
under grant PHY-0756174 and NSF PHY11-25915 and by the Department of Energy under
contract DE-AC03-76SF00515.  The research of  A.W.~was supported by the Impuls und Vernetzungsfond of the Helmholtz Association of German Research Centres under grant HZ-NG-603.
We thank the Aspen Center for Physics and the Princeton Center for Theoretical Science for hospitality during the course of this work.


\newpage



\begin{thebibliography}{99}


\bibitem{MonodromyI}

E.~Silverstein and A.~Westphal,
  ``Monodromy in the CMB: Gravity Waves and String Inflation,''
  Phys.\ Rev.\ D {\bf 78}, 106003 (2008)
  [arXiv:0803.3085 [hep-th]].
  %%CITATION = ARXIV:0803.3085;%%
  %281 citations counted in INSPIRE as of 06 Nov 2014

\bibitem{MSW}

%\cite{McAllister:2008hb}
%\bibitem{McAllister:2008hb}
  L.~McAllister, E.~Silverstein and A.~Westphal,
  ``Gravity Waves and Linear Inflation from Axion Monodromy,''
  Phys.\ Rev.\ D {\bf 82}, 046003 (2010)
  [arXiv:0808.0706 [hep-th]].
  %%CITATION = ARXIV:0808.0706;%%
  %196 citations counted in INSPIRE as of 19 Apr 2014
%\cite{Silverstein:2008sg}

\bibitem{FMPWX}

R.~Flauger, L.~McAllister, E.~Pajer, A.~Westphal and G.~Xu,
  ``Oscillations in the CMB from Axion Monodromy Inflation,''
  JCAP {\bf 1006}, 009 (2010)
  [arXiv:0907.2916 [hep-th]].
  %%CITATION = ARXIV:0907.2916;%%
  %103 citations counted in INSPIRE as of 15 Oct 2013

\bibitem{ignoble}

  N.~Kaloper and L.~Sorbo,
  ``A Natural Framework for Chaotic Inflation,''
  Phys.\ Rev.\ Lett.\  {\bf 102}, 121301 (2009)
  [arXiv:0811.1989 [hep-th]].
  %%CITATION = ARXIV:0811.1989;%%
  %71 citations counted in INSPIRE as of 14 May 2014

N.~Kaloper, A.~Lawrence and L.~Sorbo,
  ``An Ignoble Approach to Large Field Inflation,''
  JCAP {\bf 1103}, 023 (2011)
  [arXiv:1101.0026 [hep-th]].
  %%CITATION = ARXIV:1101.0026;%%
  %23 citations counted in INSPIRE as of 15 Oct 2013

N.~Kaloper and A.~Lawrence,
  ``Natural Chaotic Inflation and UV Sensitivity,''
  arXiv:1404.2912 [hep-th].
  %%CITATION = ARXIV:1404.2912;%%
  %1 citations counted in INSPIRE as of 19 Apr 2014


\bibitem{Andreichaotic}
  A.~D.~Linde,
  ``Chaotic Inflation,''
  Phys.\ Lett.\ B {\bf 129}, 177 (1983).
  %%CITATION = PHLTA,B129,177;%%
  %1664 citations counted in INSPIRE as of 11 Apr 2014

\bibitem{Natural}

K.~Freese, J.~A.~Frieman and A.~V.~Olinto,
  ``Natural inflation with pseudo - Nambu-Goldstone bosons,''
  Phys.\ Rev.\ Lett.\  {\bf 65}, 3233 (1990).
  %%CITATION = PRLTA,65,3233;%%
  %384 citations counted in INSPIRE as of 29 Oct 2013

F.~C.~Adams, J.~R.~Bond, K.~Freese, J.~A.~Frieman and A.~V.~Olinto,
  ``Natural inflation: Particle physics models, power law spectra for large scale structure, and constraints from COBE,''
  Phys.\ Rev.\ D {\bf 47}, 426 (1993)
  [hep-ph/9207245].
  %%CITATION = HEP-PH/9207245;%%
  %301 citations counted in INSPIRE as of 29 Oct 2013

\bibitem{multiax}

%\cite{Kim:2004rp}
%\bibitem{Kim:2004rp}
  J.~E.~Kim, H.~P.~Nilles and M.~Peloso,
  ``Completing natural inflation,''
  JCAP {\bf 0501}, 005 (2005)
  [hep-ph/0409138].
  %%CITATION = HEP-PH/0409138;%%
  %107 citations counted in INSPIRE as of 08 Jul 2014

%\bibitem{Nflation}

S.~Dimopoulos, S.~Kachru, J.~McGreevy and J.~G.~Wacker,
  ``N-flation,''
  JCAP {\bf 0808}, 003 (2008)
  [hep-th/0507205].
  %%CITATION = HEP-TH/0507205;%%
  %245 citations counted in INSPIRE as of 11 Oct 2013

\bibitem{Danjie}
%\cite{Wenren:2014cga}
%\bibitem{Wenren:2014cga}
  D.~Wenren,
  ``Tilt and Tensor-to-Scalar Ratio in Multifield Monodromy Inflation,''
  arXiv:1405.1411 [hep-th].
  %%CITATION = ARXIV:1405.1411;%%

\bibitem{DanGreen}
D.~Green,
  ``Disorder in the Early Universe,''
  arXiv:1409.6698 [hep-th].
  %%CITATION = ARXIV:1409.6698;%%

\bibitem{oscillationsdata}

R.~Easther and R.~Flauger,
  ``Planck Constraints on Monodromy Inflation,''
  arXiv:1308.3736 [astro-ph.CO].
  %%CITATION = ARXIV:1308.3736;%%
  %2 citations counted in INSPIRE as of 30 Oct 2013
%\cite{Peiris:2013opa}

%\bibitem{Peiris:2013opa}
  H.~Peiris, R.~Easther and R.~Flauger,
  ``Constraining Monodromy Inflation,''
  JCAP {\bf 1309}, 018 (2013)
  [arXiv:1303.2616 [astro-ph.CO]].
  %%CITATION = ARXIV:1303.2616;%%
  %16 citations counted in INSPIRE as of 30 Oct 2013

M.~G.~Jackson, B.~Wandelt and F.~Bouchet,
  ``Angular Correlation Functions for Models with Logarithmic Oscillations,''
  arXiv:1303.3499 [hep-th].
  %%CITATION = ARXIV:1303.3499;%%
  %6 citations counted in INSPIRE as of 30 Oct 2013

%\cite{Ade:2013uln}
%\bibitem{Ade:2013uln}
  P.~A.~R.~Ade {\it et al.}  [Planck Collaboration],
  ``Planck 2013 results. XXII. Constraints on inflation,''
  arXiv:1303.5082 [astro-ph.CO].
  %%CITATION = ARXIV:1303.5082;%%
  %595 citations counted in INSPIRE as of 14 May 2014


%\bibitem{Meerburg:2013cla}
  P.~D.~Meerburg, D.~N.~Spergel and B.~D.~Wandelt,
  ``Searching for Oscillations in the Primordial Power Spectrum: Perturbative Approach (Paper I),''
  Phys.\ Rev.\ D {\bf 89}, 063536 (2014)
  [arXiv:1308.3704 [astro-ph.CO]].
  %%CITATION = ARXIV:1308.3704;%%
  %6 citations counted in INSPIRE as of 14 May 2014


P.~D.~Meerburg and D.~N.~Spergel,
  ``Searching for Oscillations in the Primordial Power Spectrum: Constraints from Planck (Paper II),''
  Phys.\ Rev.\ D {\bf 89}, 063537 (2014)
  [arXiv:1308.3705 [astro-ph.CO]].
  %%CITATION = ARXIV:1308.3705;%%
  %10 citations counted in INSPIRE as of 14 May 2014
%\cite{Meerburg:2013cla}

P.~D.~Meerburg, D.~N.~Spergel and B.~D.~Wandelt,
  ``Searching for oscillations in the primordial power spectrum,''
  arXiv:1406.0548 [astro-ph.CO].
  %%CITATION = ARXIV:1406.0548;%%

P.~D.~Meerburg,
  ``Alleviating the tension at low multipole through Axion Monodromy,''
  arXiv:1406.3243 [astro-ph.CO].
  %%CITATION = ARXIV:1406.3243;%%

%\cite{Minor:2014xla}
%\bibitem{Minor:2014xla}
  Q.~E.~Minor and M.~Kaplinghat,
  ``Inflation that runs naturally: Gravitational waves and suppression of power at large and small scales,''
  arXiv:1411.0689 [astro-ph.CO].
  %%CITATION = ARXIV:1411.0689;%%




\bibitem{EFToscillations}

S.~R.~Behbahani, A.~Dymarsky, M.~Mirbabayi and L.~Senatore,
  ``(Small) Resonant non-Gaussianities: Signatures of a Discrete Shift Symmetry in the Effective Field Theory of Inflation,''
  JCAP {\bf 1212}, 036 (2012)
  [arXiv:1111.3373 [hep-th]].
  %%CITATION = ARXIV:1111.3373;%%
  %32 citations counted in INSPIRE as of 01 Nov 2014

S.~R.~Behbahani and D.~Green,
  ``Collective Symmetry Breaking and Resonant Non-Gaussianity,''
  JCAP {\bf 1211}, 056 (2012)
  [arXiv:1207.2779 [hep-th]].
  %%CITATION = ARXIV:1207.2779;%%
  %17 citations counted in INSPIRE as of 01 Nov 2014




%\cite{Dong:2010in}
\bibitem{flattening}
  X.~Dong, B.~Horn, E.~Silverstein and A.~Westphal,
  ``Simple exercises to flatten your potential,''
  Phys.\ Rev.\ D {\bf 84}, 026011 (2011)
  [arXiv:1011.4521 [hep-th]].
  %%CITATION = ARXIV:1011.4521;%%
  %47 citations counted in INSPIRE as of 03 Jun 2014


%\cite{McAllister:2014mpa}
\bibitem{powers}
  L.~McAllister, E.~Silverstein, A.~Westphal and T.~Wrase,
  ``The Powers of Monodromy,''
  arXiv:1405.3652 [hep-th].
  %%CITATION = ARXIV:1405.3652;%%
  %1 citations counted in INSPIRE as of 27 May 2014


\bibitem{recentmonodromy}

%\cite{Palti:2014kza}
%\bibitem{Palti:2014kza}
  E.~Palti and T.~Weigand,
  ``Towards large r from [p,q]-inflation,''
  arXiv:1403.7507 [hep-th].
  %%CITATION = ARXIV:1403.7507;%%
  %16 citations counted in INSPIRE as of 07 May 2014

%\cite{Marchesano:2014mla}
%\bibitem{Marchesano:2014mla}
  F.~Marchesano, G.~Shiu and A.~M.~Uranga,
  ``F-term Axion Monodromy Inflation,''
  arXiv:1404.3040 [hep-th].
  %%CITATION = ARXIV:1404.3040;%%
  %17 citations counted in INSPIRE as of 07 May 2014

%\cite{Harigaya:2014eta}
%\bibitem{Harigaya:2014eta}
  K.~Harigaya and M.~Ibe,
  ``Inflaton potential on a Riemann surface,''
  arXiv:1404.3511 [hep-ph].
  %%CITATION = ARXIV:1404.3511;%%
  %8 citations counted in INSPIRE as of 27 May 2014

%\cite{Hebecker:2014eua}
%\bibitem{Hebecker:2014eua}
  A.~Hebecker, S.~C.~Kraus and L.~T.~Witkowski,
  ``D7-Brane Chaotic Inflation,''
  arXiv:1404.3711 [hep-th].
  %%CITATION = ARXIV:1404.3711;%%
  %12 citations counted in INSPIRE as of 07 May 2014

%\cite{Ibanez:2014kia}
%\bibitem{Ibanez:2014kia}
  L.~E.~Ibanez and I.~Valenzuela,
  ``The Inflaton as a MSSM Higgs and Open String Modulus Monodromy Inflation,''
  arXiv:1404.5235 [hep-th].
  %%CITATION = ARXIV:1404.5235;%%
  %5 citations counted in INSPIRE as of 07 May 2014

%\cite{Kobayashi:2014ooa}
%\bibitem{Kobayashi:2014ooa}
  T.~Kobayashi, O.~Seto and Y.~Yamaguchi,
  ``Axion monodromy inflation with sinusoidal corrections,''
  arXiv:1404.5518 [hep-ph].
  %%CITATION = ARXIV:1404.5518;%%
  %7 citations counted in INSPIRE as of 27 May 2014

%\cite{Dine:2014hwa}
%\bibitem{Dine:2014hwa}
  M.~Dine, P.~Draper and A.~Monteux,
  ``Monodromy Inflation in SUSY QCD,''
  arXiv:1405.0068 [hep-th].
  %%CITATION = ARXIV:1405.0068;%%
  %1 citations counted in INSPIRE as of 07 May 2014

%\cite{Arends:2014qca}
%\bibitem{Arends:2014qca}
  M.~Arends, A.~Hebecker, K.~Heimpel, S.~C.~Kraus, D.~L\"ust, C.~Mayrhofer, C.~Schick and T.~Weigand,
  ``D7-Brane Moduli Space in Axion Monodromy and Fluxbrane Inflation,''
  arXiv:1405.0283 [hep-th].
  %%CITATION = ARXIV:1405.0283;%%

%\cite{Yonekura:2014oja}
%\bibitem{Yonekura:2014oja}
  K.~Yonekura,
  ``Notes on natural inflation,''
  arXiv:1405.0734 [hep-th].
  %%CITATION = ARXIV:1405.0734;%%

%\cite{Higaki:2014sja}
%\bibitem{Higaki:2014sja}
  T.~Higaki, T.~Kobayashi, O.~Seto and Y.~Yamaguchi,
  ``Axion monodromy inflation with multi-natural modulations,''
  arXiv:1405.0775 [hep-ph].
  %%CITATION = ARXIV:1405.0775;%%


%\cite{Hebecker:2014kva}
%\bibitem{Hebecker:2014kva}
  A.~Hebecker, P.~Mangat, F.~Rompineve and L.~T.~Witkowski,
  ``Tuning and Backreaction in F-term Axion Monodromy Inflation,''
  arXiv:1411.2032 [hep-th].
  %%CITATION = ARXIV:1411.2032;%%

%\cite{Carone:2014cta}
%\bibitem{Carone:2014cta}
  C.~D.~Carone, J.~Erlich, A.~Sensharma and Z.~Wang,
  ``Dante's Waterfall,''
  arXiv:1410.2593 [hep-ph].
  %%CITATION = ARXIV:1410.2593;%%

%\cite{Blumenhagen:2014nba}
%\bibitem{Blumenhagen:2014nba}
  R.~Blumenhagen, D.~Herschmann and E.~Plauschinn,
  ``The Challenge of Realizing F-term Axion Monodromy Inflation in String Theory,''
  arXiv:1409.7075 [hep-th].
  %%CITATION = ARXIV:1409.7075;%%
  %2 citations counted in INSPIRE as of 13 Nov 2014


%\cite{Franco:2014hsa}
 %\bibitem{Franco:2014hsa}
  S.~Franco, D.~Galloni, A.~Retolaza and A.~Uranga,
  ``Axion Monodromy Inflation on Warped Throats,''
  arXiv:1405.7044 [hep-th].
  %%CITATION = ARXIV:1405.7044;%%
  %5 citations counted in INSPIRE as of 13 Nov 2014

L.~E.~Ibanez, F.~Marchesano and I.~Valenzuela,
  ``Higgs-otic Inflation and String Theory,''
  arXiv:1411.5380 [hep-th].
  %%CITATION = ARXIV:1411.5380;%%




\bibitem{SC}

M.~Dodelson, X.~Dong, E.~Silverstein and G.~Torroba,
  ``New solutions with accelerated expansion in string theory,''
  arXiv:1310.5297 [hep-th];
  %%CITATION = ARXIV:1310.5297;%%
  %6 citations counted in INSPIRE as of 02 Nov 2014

and additional work in progress with Gur-Ari, Wenren, et al.

\bibitem{twistedtori}

G.~Gur-Ari,
  ``Brane Inflation and Moduli Stabilization on Twisted Tori,''
  JHEP {\bf 1401}, 179 (2014)
  [arXiv:1310.6787 [hep-th]].
  %%CITATION = ARXIV:1310.6787;%%
  %8 citations counted in INSPIRE as of 06 Nov 2014

\bibitem{unwinding}
G.~D'Amico, R.~Gobbetti, M.~Kleban and M.~Schillo,
  ``Unwinding Inflation,''
  JCAP {\bf 1303}, 004 (2013)
  [arXiv:1211.4589 [hep-th]].
  %%CITATION = ARXIV:1211.4589;%%
  %19 citations counted in INSPIRE as of 06 Nov 2014

\bibitem{LesHouches}
 E.~Silverstein,
  ``Les Houches lectures on inflationary observables and string theory,''
  arXiv:1311.2312 [hep-th].
  %%CITATION = ARXIV:1311.2312;%%
  %8 citations counted in INSPIRE as of 06 Nov 2014


\bibitem{KKLT}
S.~Kachru, R.~Kallosh, A.~D.~Linde and S.~P.~Trivedi,
  ``De Sitter vacua in string theory,''
  Phys.\ Rev.\ D {\bf 68}, 046005 (2003)
  [hep-th/0301240].
  %%CITATION = HEP-TH/0301240;%%
  %2015 citations counted in INSPIRE as of 03 Nov 2014

\bibitem{LARGE}

V.~Balasubramanian, P.~Berglund, J.~P.~Conlon and F.~Quevedo,
  ``Systematics of moduli stabilisation in Calabi-Yau flux compactifications,''
  JHEP {\bf 0503}, 007 (2005)
  [hep-th/0502058].
  %%CITATION = HEP-TH/0502058;%%
  %474 citations counted in INSPIRE as of 03 Nov 2014
 %
%  %326 citations counted in INSPIRE as of 05 Nov 2014


\bibitem{worldsheetinstantons}
 M.~Dine, N.~Seiberg, X.~G.~Wen and E.~Witten,
  ``Nonperturbative Effects on the String World Sheet. 2.,''
  Nucl.\ Phys.\ B {\bf 289}, 319 (1987).
  %%CITATION = NUPHA,B289,319;%%
  %275 citations counted in INSPIRE as of 06 Nov 2014
%\cite{Dine:1986zy}

%\bibitem{Dine:1986zy}
  M.~Dine, N.~Seiberg, X.~G.~Wen and E.~Witten,
  ``Nonperturbative Effects on the String World Sheet,''
  Nucl.\ Phys.\ B {\bf 278}, 769 (1986).
  %%CITATION = NUPHA,B278,769;%%


\bibitem{LiamDanielBook}

D.~Baumann and L.~McAllister,
  ``Inflation and String Theory,''
  arXiv:1404.2601 [hep-th].
  %%CITATION = ARXIV:1404.2601;%%
  %51 citations counted in INSPIRE as of 12 Nov 2014

%\cite{Feroz:2007kg}
\bibitem{multinest}
  F.~Feroz and M.~P.~Hobson,
  ``Multimodal nested sampling: an efficient and robust alternative to MCMC methods for astronomical data analysis,''
  Mon.\ Not.\ Roy.\ Astron.\ Soc.\  {\bf 384}, 449 (2008)
  [arXiv:0704.3704 [astro-ph]].
  %%CITATION = ARXIV:0704.3704;%%
  %245 citations counted in INSPIRE as of 01 Dec 2014

%\cite{Feroz:2008xx}
%\bibitem{Feroz:2008xx}
  F.~Feroz, M.~P.~Hobson and M.~Bridges,
  ``MultiNest: an efficient and robust Bayesian inference tool for cosmology and particle physics,''
  Mon.\ Not.\ Roy.\ Astron.\ Soc.\  {\bf 398}, 1601 (2009)
  [arXiv:0809.3437 [astro-ph]].
  %%CITATION = ARXIV:0809.3437;%%
  %323 citations counted in INSPIRE as of 01 Dec 2014

%\cite{Feroz:2013hea}
%\bibitem{Feroz:2013hea}
  F.~Feroz, M.~P.~Hobson, E.~Cameron and A.~N.~Pettitt,
  ``Importance Nested Sampling and the MultiNest Algorithm,''
  arXiv:1306.2144 [astro-ph.IM].
  %%CITATION = ARXIV:1306.2144;%%
  %36 citations counted in INSPIRE as of 01 Dec 2014

%\cite{Flauger:2010ja}
\bibitem{Flauger:2010ja}
  R.~Flauger and E.~Pajer,
  %``Resonant Non-Gaussianity,''
  JCAP {\bf 1101}, 017 (2011)
  [arXiv:1002.0833 [hep-th]].
  %%CITATION = ARXIV:1002.0833;%%
  %97 citations counted in INSPIRE as of 03 Dec 2014

%\cite{Ade:2013kta}
\bibitem{Ade:2013kta}
  P.~A.~R.~Ade {\it et al.}  [Planck Collaboration],
  ``Planck 2013 results. XV. CMB power spectra and likelihood,''
  Astron.\ Astrophys.\  {\bf 571}, A15 (2014)
  [arXiv:1303.5075 [astro-ph.CO]].
  %%CITATION = ARXIV:1303.5075;%%
  %300 citations counted in INSPIRE as of 25 Nov 2014

%\cite{Ade:2013uln}
\bibitem{PlanckInf}
  P.~A.~R.~Ade {\it et al.}  [Planck Collaboration],
  ``Planck 2013 results. XXII. Constraints on inflation,''
  Astron.\ Astrophys.\  {\bf 571}, A22 (2014)
  [arXiv:1303.5082 [astro-ph.CO]].
  %%CITATION = ARXIV:1303.5082;%%
  %858 citations counted in INSPIRE as of 25 Nov 2014

%\cite{Spergel:2013rxa}
\bibitem{Spergel:2013rxa}
  D.~Spergel, R.~Flauger and R.~Hlozek,
  ``Planck Data Reconsidered,''
  arXiv:1312.3313 [astro-ph.CO].
  %%CITATION = ARXIV:1312.3313;%%
  %46 citations counted in INSPIRE as of 25 Nov 2014


 
\bibitem{RFCosmo}
  R.~Flauger, talk at COSMO 2013.
  
\end{thebibliography}
\end{document}